\documentclass[nologo,11pt,a4paper]{ETHpaper}
\usepackage{graphicx, amsmath, amssymb,color,wasysym}
\usepackage[square,numbers,sort&compress]{natbib}
\usepackage{calc}
\usepackage{caption}
\usepackage{subcaption}
\usepackage{longtable}
\usepackage{booktabs}
\usepackage{mathtools}
\usepackage{cancel,cool}
\usepackage{cleveref}
\usepackage{bm}

\title{Probing the robustness of nested multi-layer networks}
\author{Giona Casiraghi, Antonios Garas, Frank Schweitzer}
\address{ETH Z\"urich, Chair of Systems Design\\ Weinbergstrasse 56/58,  Z\"urich, Switzerland}
\www{\url{http://www.sg.ethz.ch}}
\reference{(Submitted for publication)} 

\begin{document}

\makeframing
\maketitle

\begin{abstract}
  We consider a multi-layer network with two layers, $\mathcal{L}_{1}$, $\mathcal{L}_{2}$. 
  Their intra-layer topology shows a scale-free degree distribution and a core-periphery structure.
  A nested structure describes the inter-layer topology, i.e., some nodes from $\mathcal{L}_{1}$, the generalists, have many links to nodes in $\mathcal{L}_{2}$, specialists only have a few.
  This structure is verified by analyzing two empirical networks from ecology and economics.
  To probe the robustness of the multi-layer network, we remove nodes from $\mathcal{L}_{1}$ with their inter- and intra-layer links and measure the impact on the size of the largest connected component, $F_{2}$, in $\mathcal{L}_{2}$, which we take as a robustness measure. 
  We test different attack scenarios by preferably removing peripheral or core nodes.
  We also vary the intra-layer coupling between generalists and specialists, to study their impact on the robustness of the multi-layer network. 
  We find that some combinations of attack scenario and intra-layer coupling lead to very low robustness values, whereas others demonstrate high robustness of the multi-layer network because of the intra-layer links. 
  Our results shed new light on the robustness of bipartite networks, which consider only inter-layer, but no intra-layer links.

    \emph{Keywords: bipartite network, nestedness, attack scenario, robustness profile}
  \end{abstract}
\date{\today}

\section{Introduction}
\label{sec:introduction}

Heterogeneity matters -- also in complex networks.
Nodes in real-world networks are most often \emph{not similar}; they are very \emph{different} with respect to their \emph{degree}, i.e., the number of links to connect them to other nodes. 
The concept of \emph{scale-free networks}~\citep{Barabasi1999} reflects the fact that the degree distribution is not only very broad, it often also lacks a characteristic scale to measure degrees~\citep{Schweitzer2017}.
Even nodes with a \emph{very large} degree have a considerable probability to appear.
Whether they are indeed observed in real networks is mostly constrained by the size and the quality of the data sets, which appears to be a bigger problem.
But also the statistical methods used to estimate these probabilities play a role in the current discussion about scale-free networks~\citep{broido2019scale}.

Scale-free networks became of interest because of their \emph{robust, yet fragile} structure in case of attacks.
Nodes with a large degree, called \emph{hubs}, are rare, nodes with a small degree very frequent.
An attack that \emph{randomly} chooses a node for removal will therefore most often hit nodes with low degree.
Their removal does not change much topological properties of the network, such as connectedness.
However, if the attack is targeted toward the hubs, this considerably impacts the topology.
Thus, it was argued that the robustness of scale-free networks against attacks could be much improved if these hubs are protected.
We will come back to this topological argumentation later in our paper.
At this point, we just remark that the perspective may change if a dynamics \emph{on} the network is considered \citep{burkholz2016damage}.

In recent years, the investigation of scale-free networks was extended from \emph{single-layer} to \emph{multi-layer} networks~\cite{Garas2016,kivela2014multilayer,Wider2016}.
This development was driven by the insight that systems can be hardly studied in isolation.
Taking the example of a communication network, on the one hand, and the power grid, on the other hand, it becomes obvious that these two networks mutually influence each other~\citep{Huang2011}.

To study such multi-layer networks requires to solve a number of problems, both methodological and practical ones.
To what extent can insights from single-layer networks be generalised to multi-layer networks?
For example, peripheral nodes, considered unimportant in a single-layer network, can play an important role when coupling different layers~\citep{Zhang2016}.
The existence of inter-layer couplings also challenge the robustness of such \emph{networks of networks}~\citep{Gao2011}, because they allow for failure propagation between different layers~\citep{Burkholz2016}. 

On the empirical side, the biggest problem is the availability of data for \emph{both} inter-layer and \emph{intra-layer} interactions, which would allow reconstructing a multi-layer network. 
Such data is, unfortunately, only available in rare cases~\cite{Kefi2015}.
Therefore, instead of a complete multi-layer network, most of the time, only its projection as a \emph{bipartite} network is studied.
Bipartite networks neglect \emph{intra-layer} links, to only focus on the \emph{inter-layer} interaction.
Ecological networks, for instance, are most often represented as bipartite networks, reflecting mutualistic interactions e.g., between plant species and pollinator species~\citep{Memmott1999,Memmott2004a} (see also Section~\ref{sec:intra-layer-degree}).
The methodology for studying ecological network was also adopted for economic and social networks~\cite{May2008,Garas2018}.

This raised the question of to what degree insights from bipartite networks or from their single-layer projections remain valid for the bigger picture, the multi-layer networks.
This regards particularly estimates about the \emph{robustness} of these networks. 
Studies have shown that the inter-layer network structure affects the stability of ecological communities~\cite{Bascompte2007, Saavedra:2011,Allesina2009}.
In particular, it was found that a particular topology of the bipartite network, the \emph{nested structure}, improves the robustness of an ecosystem~\cite{Morris2003,Memmott2004a,Mariani2019}.  

Recently, extinction patterns have been investigated even in multi-layer ecological networks ~\citep{Pilosof2017}.
But most of the time, the robustness of ecological networks is estimated from this bipartite representation.
The error made by neglecting intra-layer interactions, e.g., between plants, or between pollinators is not known.
Therefore, in this paper, we contribute to answering this question by studying a \emph{synthetic multi-layer network}, as explained in Section \ref{sec:multilayer-network}.
In the absence of reliable data about multi-layer networks, this approach allows us to systematically investigate the impact of different inter-layer couplings together with different intra-layer topologies.
For the latter, we consider a \emph{scale-free degree distribution}, but additionally a \emph{core-periphery structure} as observed in many real-world networks (see Section~\ref{sec:inter-layer-degree}).
To specify the intra-layer topology, we first study two empirical networks, from ecology and from economics (see Section~\ref{sec:intra-layer-degree}).
We further introduce different attack strategies to probe the robustness of the synthetic multi-layer network, as outlined in Section~\ref{sec:scen-probe-robustn}.
Our results both for the ecological network and for the systematic study of the synthetic network are presented in Section~\ref{result}.
We conclude in Section~\ref{discussion}, linking our investigations back to topics like systemic risk and network controllability.

\section{Methods}\label{mat-meth}

\subsection{A multi-layer network}
\label{sec:multilayer-network}

In the following we consider interconnected networks~\citep{Garas2016}.
These networks consist of \emph{layers}, which usually have different topologies.
If the nodes in each layer are the same, these networks are called \emph{multiplex} networks.
The links in each layer then represent different types of interactions between the same nodes.
By taking the example of an economic multiplex network~\citep{Schweitzer2009}, the nodes can represent firms and the links in layer 1 \emph{ownership} relations, while the links in layer 2 can represent \emph{knowledge exchange}, the links in layer 3 \emph{credit} relations, or reputation spillovers, etc.~\citep{Zhang2019}.

\begin{figure}[htbp]
  \centering
  \quad\hfill\begin{subfigure}[b]{0.52\textwidth}
  \includegraphics[width=\textwidth]{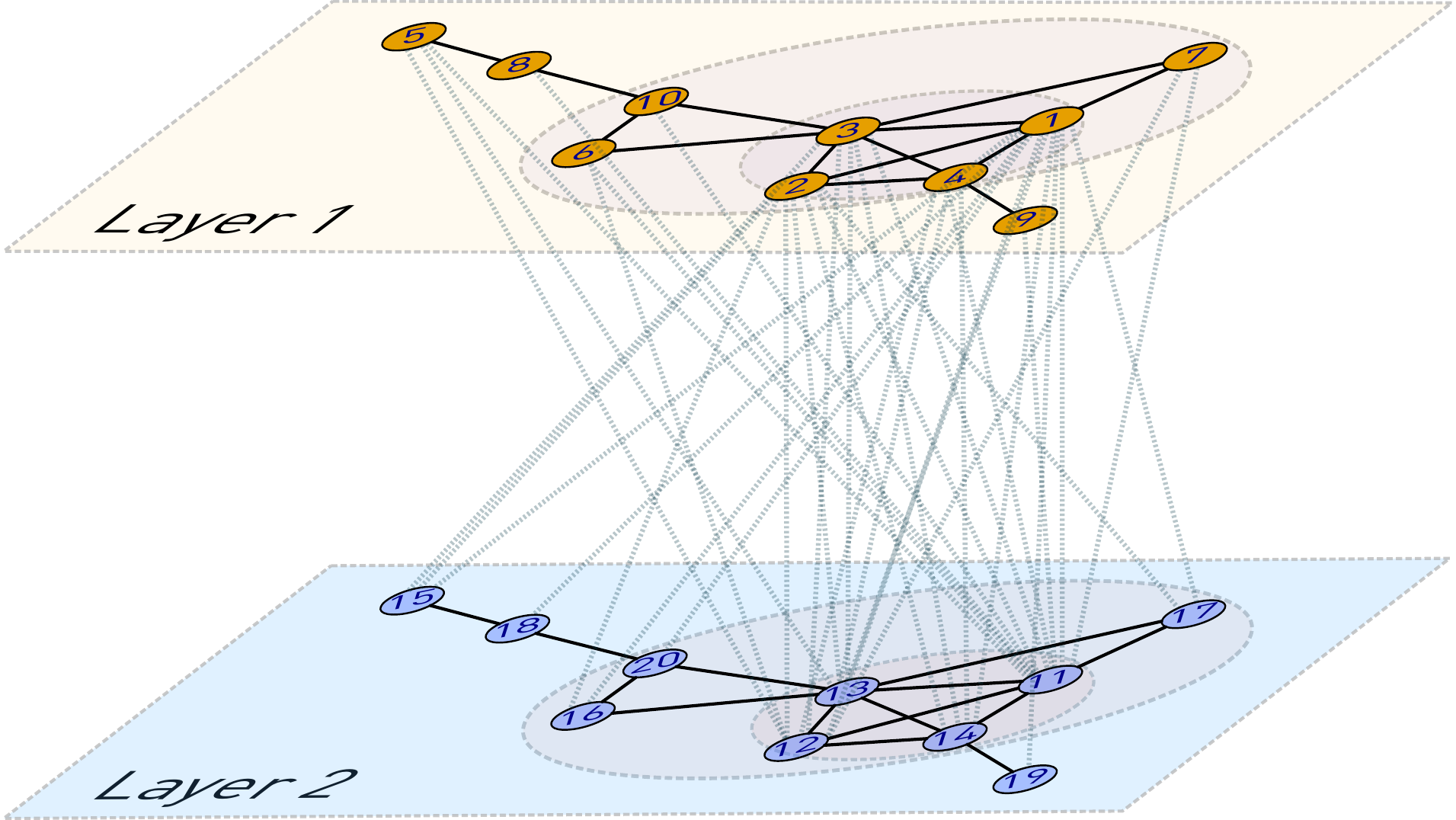}
  \caption{}
  \end{subfigure}\hfill\quad\hfill
  \begin{subfigure}[b]{0.3\textwidth}
  \includegraphics[width=\textwidth]{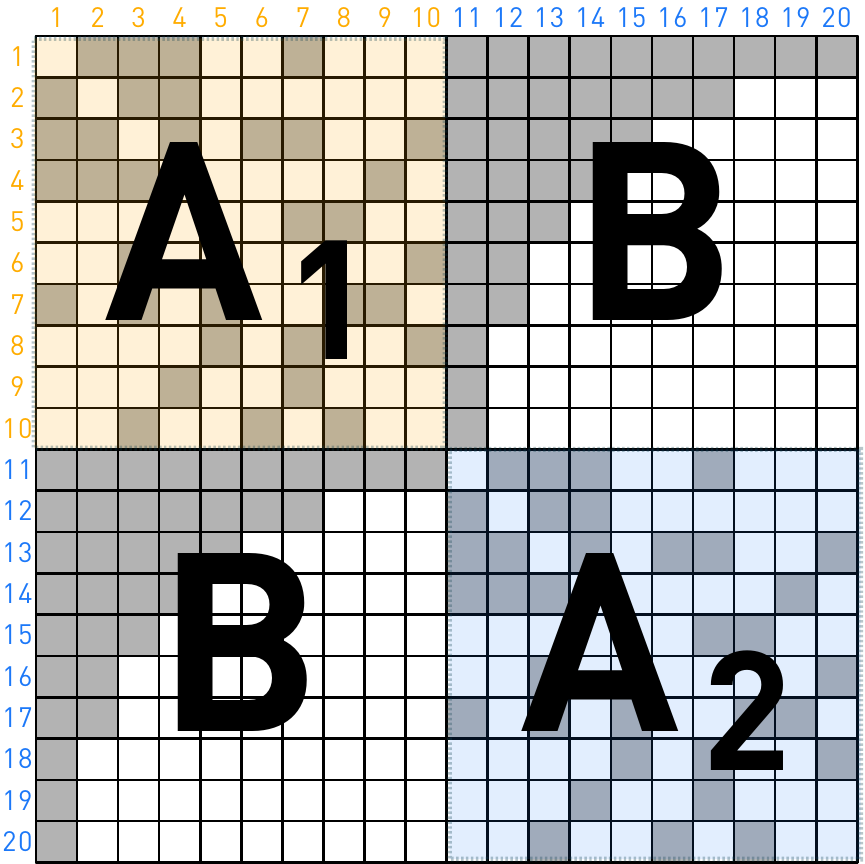}
  \caption{}
  \end{subfigure}\hfill\quad
    \caption[]{
      \textbf{(a)}   Illustration of an interconnected network with two layers. Each layer contains 10 nodes and shows a core-periphery structure. \textbf{(b)}  Supra adjacency matrix $\mathbf{A}$ of the interconnected network shown in (a).
      $\bm A_{1}$ and $\bm A_{2}$ capture the \emph{inter-layer} links in layers 1 and 2, while $\bm B$ captures the \emph{intra-layer} links between layers 1 and 2.  
  }
  \label{fig:exmulti-3}
\end{figure}

In the more general case, denoted as \emph{multi-layer} network, the nodes in different layers are also different.
The focus then is mostly on the coupling between nodes in different layers. 
Widely known are \emph{bipartite networks}, e.g., ecological networks between plants, in layer 1, and pollinators, in layer 2, as further discussed in Section~\ref{sec:intra-layer-degree}. 
Bipartite networks usually neglect links \emph{within} each layer, to address the coupling between layers.
Multi-layer networks are more general in this respect, as they allow to consider \emph{both} intra-layer and inter-layer links, at the same time.
An illustration of a \emph{two-layer network} is shown in Figure~\ref{fig:exmulti-3}.

We can now characterise each of the nodes in a multi-layer network.
Because the nodes in each layer are different anyway, we just enumerate all the nodes consecutively, regardless of their layer, i.e., $i=1,\dots,N$.
While we use $N=20$ for the illustrations, the simulations on the networks are done with $N=1.000$.
In both cases, each layer contains $N/2$ nodes. 
All possible combinations to link these different nodes are then considered in an \emph{adjacency matrix} $\mathbf A\in\mathbb N^{N\times N}$ in which the elements $a_{ij}$ are either $0$ or $1$.
This is illustrated in Figure~\ref{fig:exmulti-3}(b).
We call $\mathbf{A}$ the \emph{supra adjacency matrix} because it contains both the information about inter-layer and intra-layer links.
Specifically, $\mathbf{A}$ is composed of three different sub-matrices: $\mathbf{A}_{1}$ only contains the \emph{inter-layer} links from layer 1, while $\mathbf{A}_{2}$ contains only those from layer 2.
$\mathbf{B}$, on the other hand, contains all information about the links \emph{between} layers 1 and 2.
Because here we consider undirected links, there is only one sub-matrix $\mathbf{B}$.

The sub-matrices allow us to define for each node two different degrees, the \emph{intra-layer degree}, $l_{i}$, and the \emph{inter-layer degree}, $d_{i}$, which are discussed separately, in the following.

\subsection{Intra-layer topology}
\label{sec:inter-layer-degree}

\paragraph{Degree distribution. \ }

For the link distribution \emph{within} a given layer, we  assume a \emph{scale-free} distribution: $P(l)\propto l^{-\gamma}$ with exponent $\gamma \simeq 2.3$.
Note that the intra-layer degrees of nodes in both layers are drawn from the same distribution.
To further specify the intra-layer topology, we consider a \emph{core-periphery} structure of the networks~\cite{rombach2014core}, which is found in many real-world networks~\citep{snyder1979structural,nemeth1985international,smith1992structure,Wesley2014}. 
It means that in both layers, the core nodes have many connections to other core nodes, while nodes in the periphery are only loosely connected to the core and to other peripheral nodes.

\paragraph{Coreness. \ }

The embeddedness of nodes in a core-pheriphery network can be quantified by means of the \emph{$k$-core decomposition}~\citep{Dorogovtsev2006} which assigns a \emph{coreness value} $k_{i}$ to each node. 
The decomposition removes, from a given network, all nodes with a degree $l_{i}\leq k$ recursively, until only nodes with degree $k+1$ are left.
The procedure starts with $k=1$ and stops when all nodes left have a degree $l_{i} \leq k_{\mathrm{max}}$.
The corresponding $k$-\emph{shell} then consists of all nodes that are in a $k$-core but not in the
$(k+1)$-core, i.e.,  nodes assigned to a $k$-shell have a coreness value $k_{i}=k$.

\begin{figure}[htbp]
\centering
\quad\hfill\begin{subfigure}[b]{.4\textwidth}
  \includegraphics[width=\textwidth]{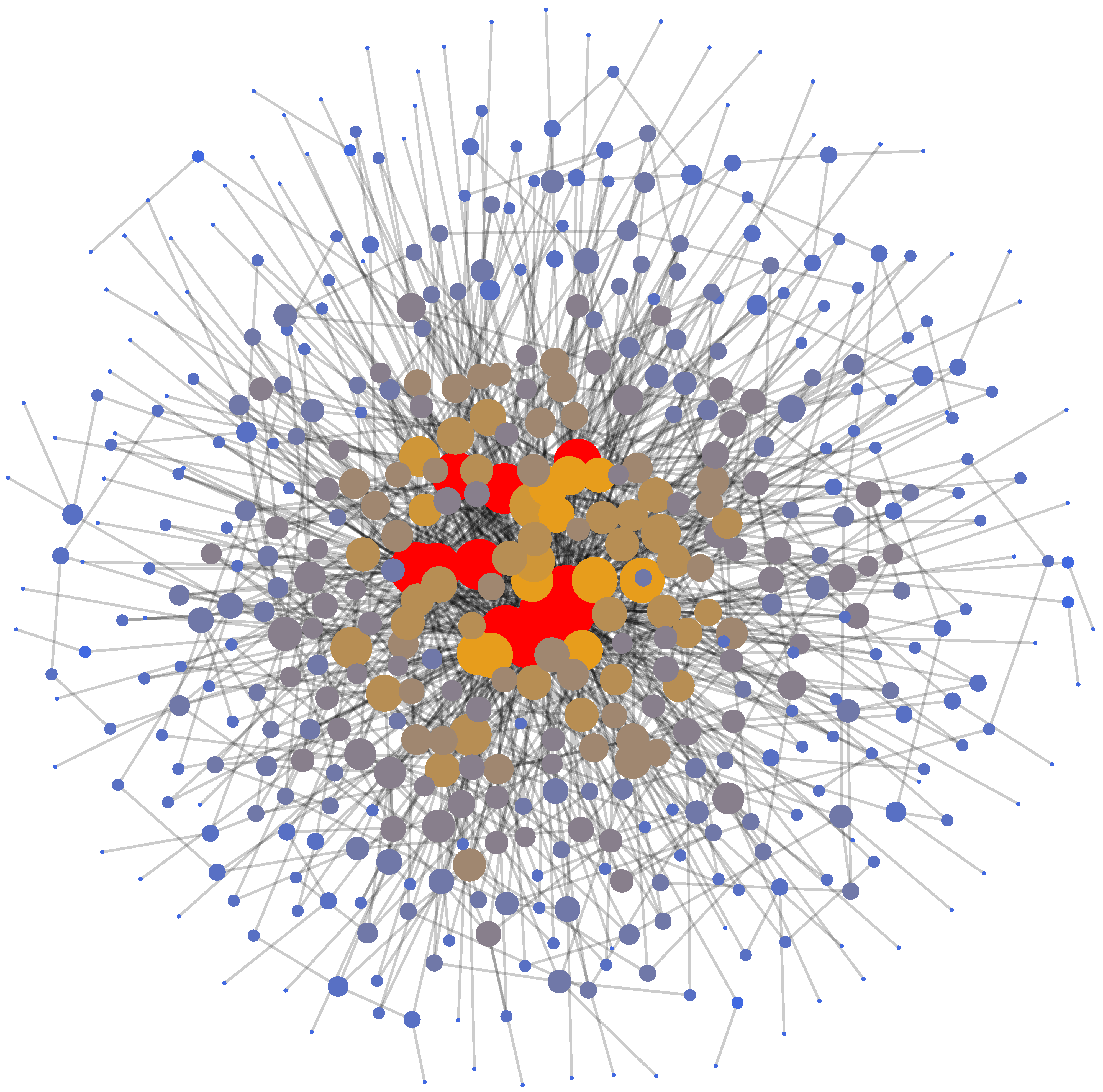}
  \caption{}
  \end{subfigure}\hfill\quad\hfill
  \begin{subfigure}[b]{.45\textwidth}
\includegraphics[width=\textwidth]{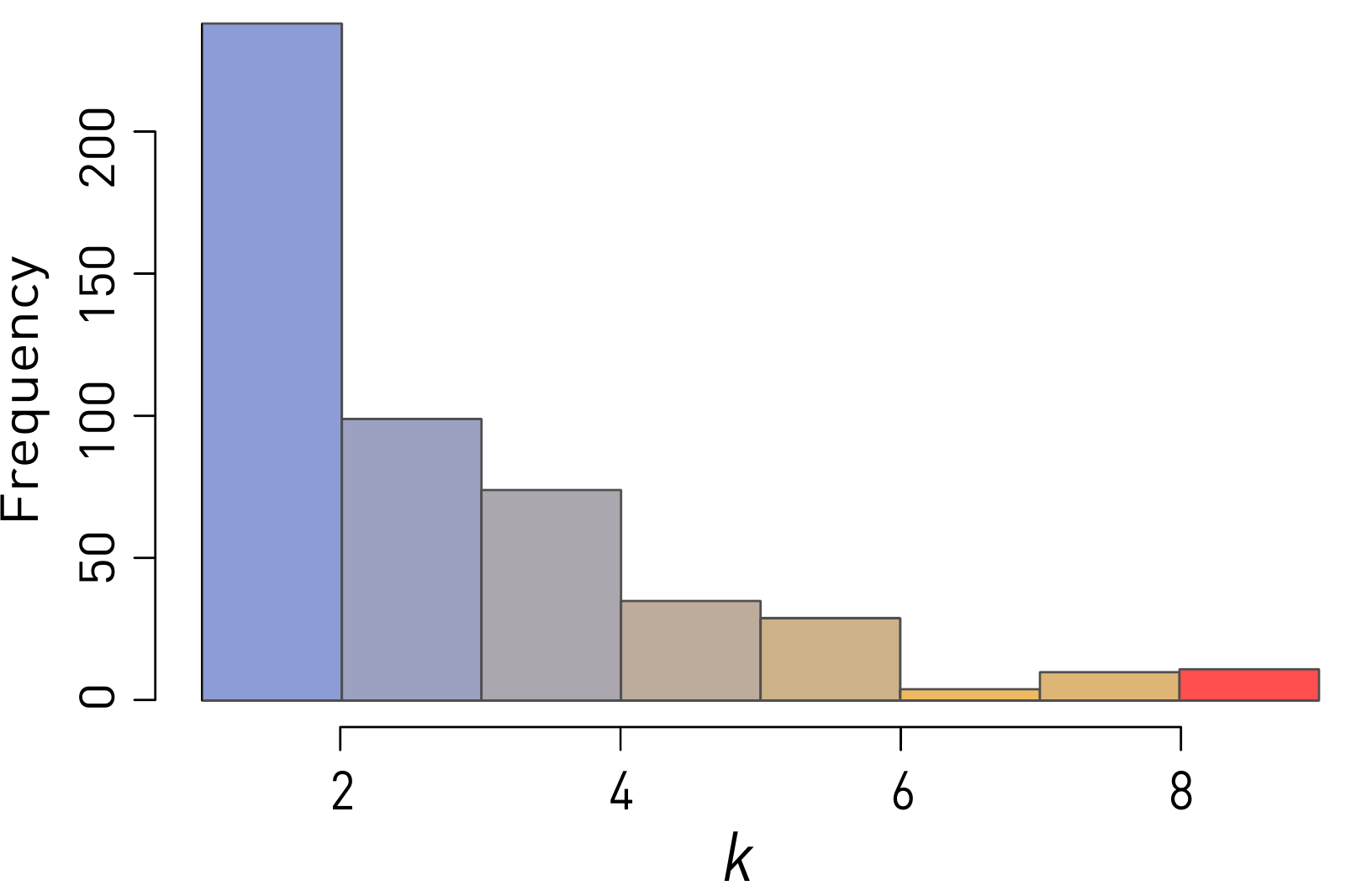}
\caption{}
  \end{subfigure}\hfill\quad
  \caption[]{
   {\bf (a)} Core-periphery topology of a network layer with 500 nodes.
    The size of a node is proportional to its degree, and its color indicates its coreness value $k_{i}$.
    {\bf (b)} Histogram of the coreness values $k$ of the network shown in (a). The colors correspond to the network.
     }
  \label{fig:exmulti-core}
\end{figure}

Figure~\ref{fig:exmulti-core}(a) provides an illustration of the core-periphery structure of each network layer, while Figure~\ref{fig:exmulti-core}(b) shows the corresponding distribution of coreness values, $P(k)$.
Nodes with low coreness values, $k_{i}\to 1$, are located in the \emph{periphery}, i.e., they are loosely connected to the core, even if some of them have a relatively high degree. 
Nodes with high coreness values, $k_{i}\to k_{\mathrm{max}}$, are densely connected and belong to the core (indicated by red colour). 
We note that, as in many real networks ~\citep{tomasello2016rise,Garas2012c,Bascompte2007}, the majority of the nodes are located in the periphery, and only a small number of nodes belong to the core.

\subsection{Inter-layer topology}
\label{sec:intra-layer-degree}

To complete the multi-layered network, we now need to specify the inter-layer topology. 
For the \emph{intra-layer} links, we used empirical findings to motivate the core-periphery structure.
Thus, for the \emph{inter-layer} link structure, we also seek inspiration from real-world networks.

\paragraph{Bipartite networks. \ }

\begin{figure}[htbp]
\centering
  \quad\hfill\begin{subfigure}[b]{0.45\textwidth}
  \includegraphics[width=\textwidth]{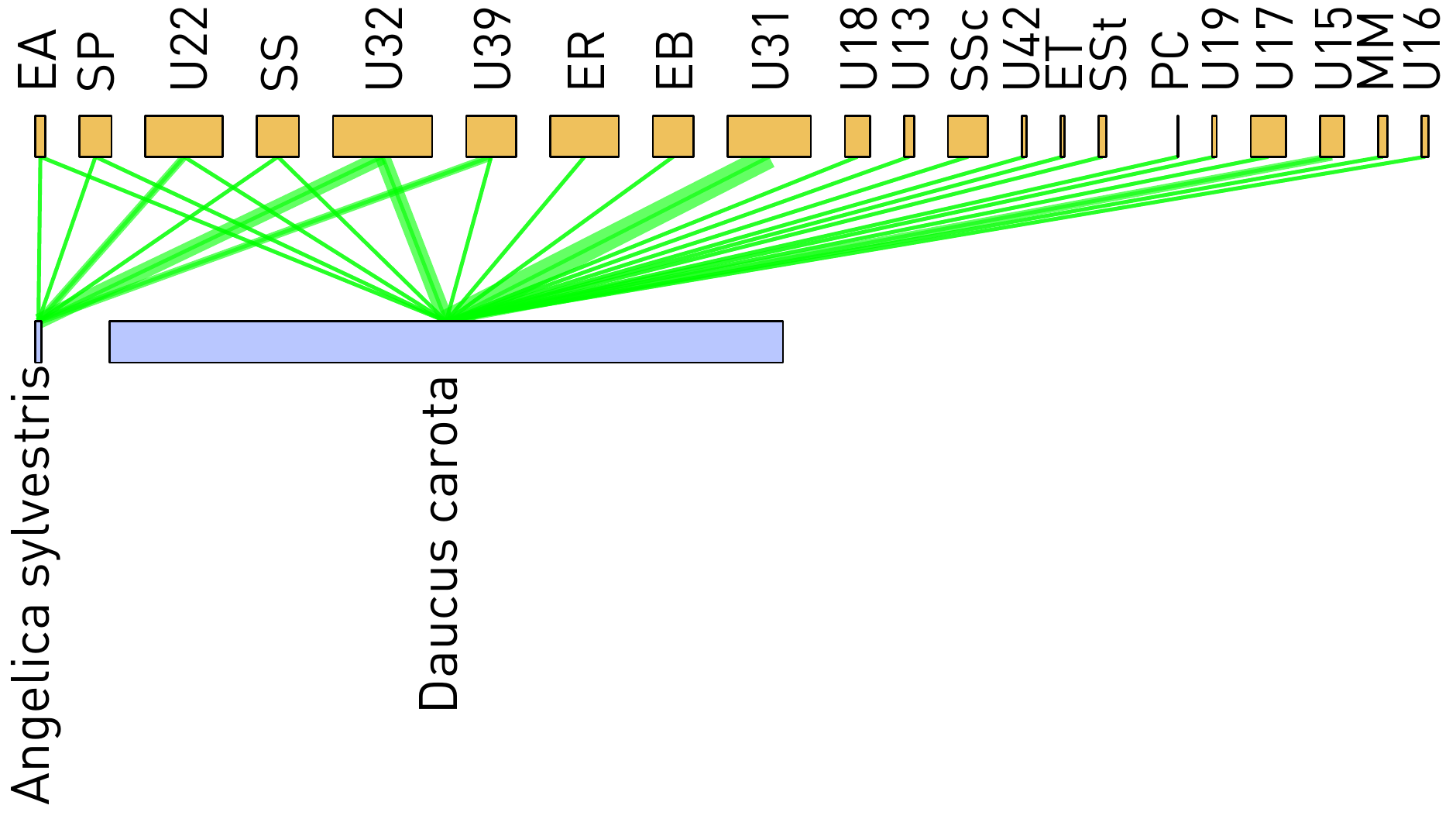}
\caption{}
  \end{subfigure}
  \hfill\quad\hfill\begin{subfigure}[b]{.45\textwidth}
  \includegraphics[width=\textwidth]{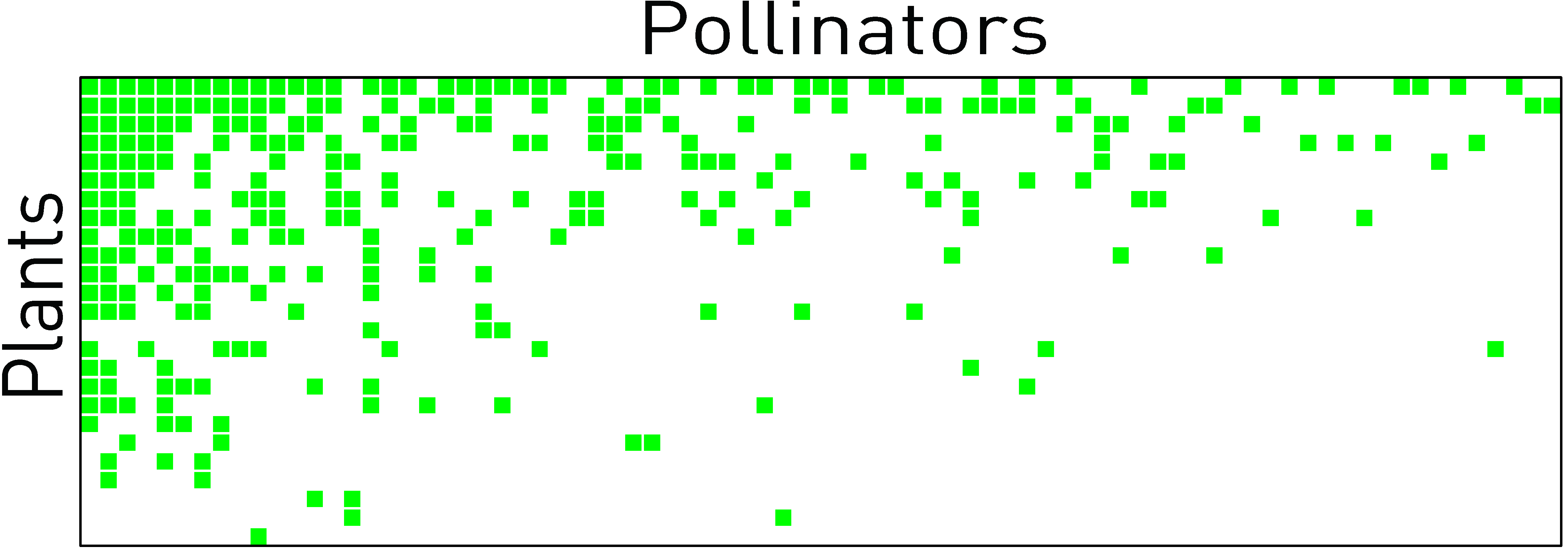}\caption{}
  \end{subfigure}\hfill\quad

  \bigskip
  
  \quad\hfill\begin{subfigure}[b]{0.45\textwidth}
  \includegraphics[width=\textwidth]{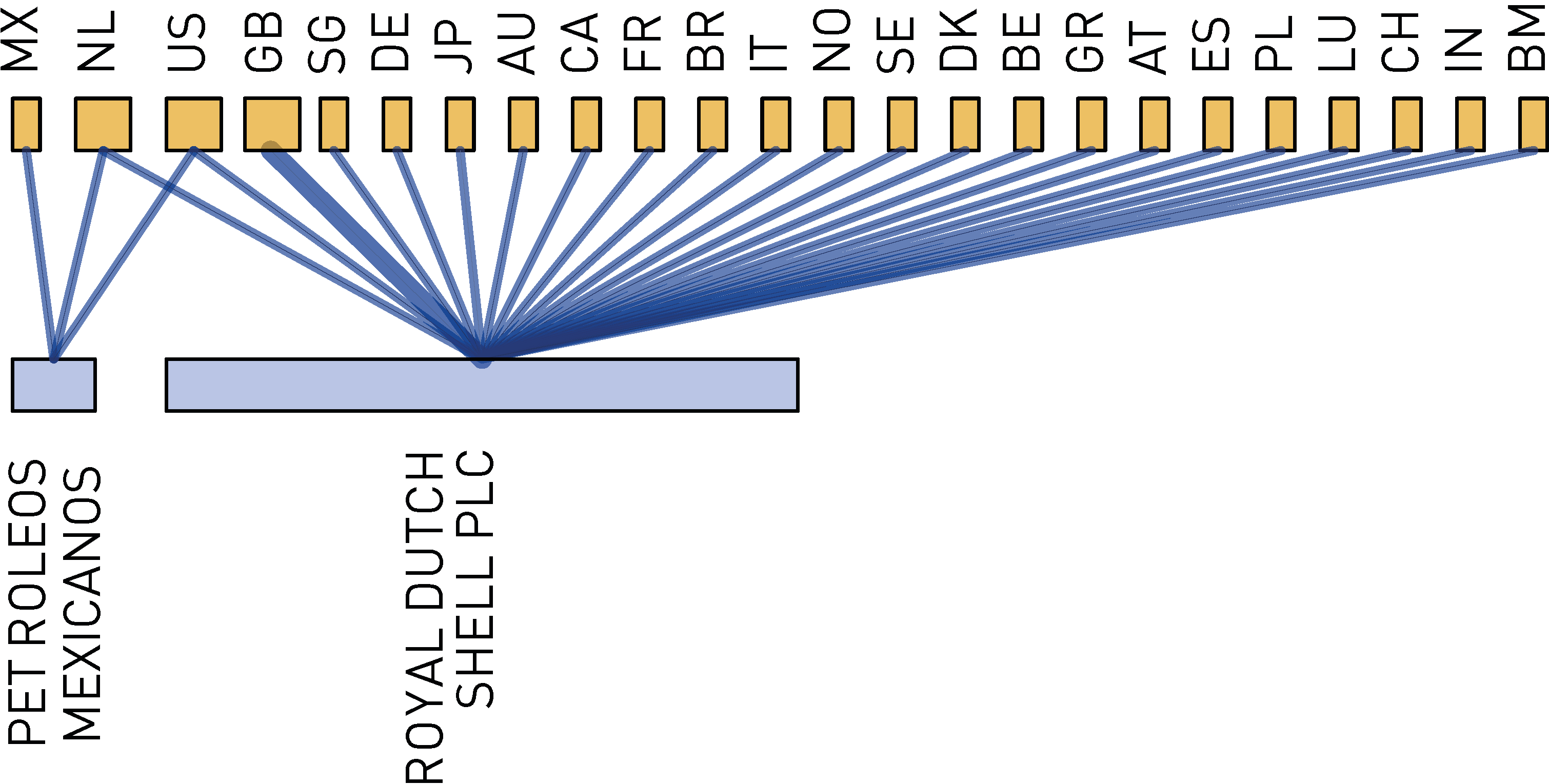}
  \caption{}
  \end{subfigure}
  \hfill\quad\hfill\begin{subfigure}[b]{.45\textwidth}
  \includegraphics[width=\textwidth]{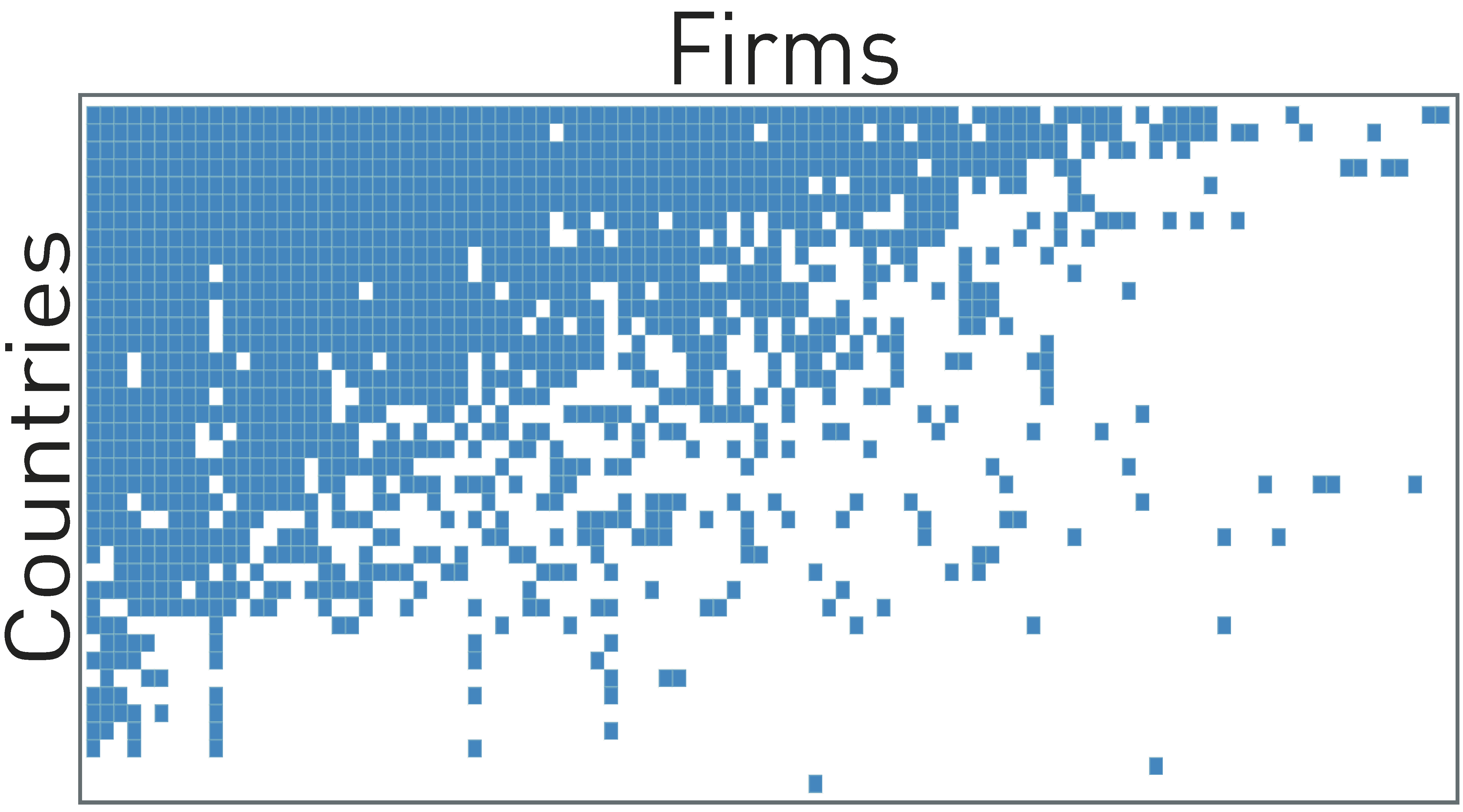}
  \caption{}
  \end{subfigure}\hfill\quad
\caption[]{ {\bf (a,c)} Partial view of a bipartite network which links (a) two plants with their pollinators and (b) two firms with the countries their subsidiaries are located in.
    {\bf (b,d)} Incidence matrices for (b) the ecological and (c) the economic bipartite network.
      The matrices are sorted using the NODF2 nestedness algorithm~\cite{Almeida-Neto:2008}.
  }
  \label{fig:bio-bip}
\end{figure}

We already pointed out that, in a multi-layer network, nodes in different layers represent different types of entities, very similar to a bipartite network with two layers $\mathcal{L}_{1}$ and $\mathcal{L}_{2}$. 
Let us consider an example from \emph{ecology} where nodes in $\mathcal{L}_{1}$ represent 25 plant \emph{species} (not individuals) and nodes in $\mathcal{L}_{2}$  79 pollinator species~\cite{Memmott1999}.
Links between these different types of nodes then represent observed  \emph{mutualistic interactions}.
Figure~\ref{fig:bio-bip}(a) shows a partial view for only two plant species. 
The thickness of the links encodes the number of observations, i.e., how often individuals of a given pollinator species have visited a plant of this species during a fixed time interval. 
The size of the plant nodes is proportional to the number of pollinator species for each plant species, and the size of the pollinator nodes is proportional to the number of plant species that are served by this pollinator species.
Taking all plant species into account, we obtain the bipartite network shown in Figure~\ref{fig:bio-bip}(b) by means of its connectivity matrix, also called \emph{incidence matrix}.
This neglects any weights, i.e.
filled/empty matrix elements indicate just the presence or absence of links in the bipartite network.
We recall that this matrix is denoted as $\mathbf{B}$ in the supra adjacency matrix shown in Figure~\ref{fig:exmulti-3}(b) because it refers to the \emph{inter-layer links}. 

To highlight the generality of such examples across disciplines, we also consider a bipartite network from \emph{economics} that links 100 large multinational firms to the 39 different countries in which their subsidiaries are located~\cite{Garas2018}.
A partial view of this network is shown in Figure~\ref{fig:bio-bip}(c), where link weights encode how many subsidiaries a firm has in a given country.
The size of the firm-nodes is proportional to the number of subsidiaries of each firm, and the size of the country-nodes is proportional to the number of subsidiaries that are present in this country.
The unweighted incidence matrix is shown in  Figure~\ref{fig:bio-bip}(d).

From the bipartite network, one can obtain two projections into single-layer networks.
Taking the ecological example, we could assume that two plant species are linked if they share the same pollinator species, or two pollinator species are linked if they serve the same plant species.
A simple exercise shows that these single-layer projections result in rather densely connected networks.
Hence, these \emph{inferred interactions} may \emph{not} reflect the real \emph{intra-layer} interactions among pollinators or among plants.
Features of the intra-layer topology, such as a scale-free distribution and a core-periphery structure, therefore cannot simply be explained from a projection of a bipartite structure, they have their independent origin~\citep{Kefi2015,Pilosof2017}.

\paragraph{Nestedness. \ }

To further characterise the structure of these two bipartite networks, we can evaluate the network density by calculating how filled the incidence matrix is.
For the ecological network, the fill is 15.1\% and for the economic network 36.6\%, which means that only 15.1\% (36.6\% respectively) of all possible inter-layer links have been observed.

As shown in Figure~\ref{fig:bio-bip}(b,d), the links of both bipartite networks are organised in a so-called \emph{nested} structure.
For the economic network, this means that only some firms are present in a large number of countries, while most other firms are present only in a few countries.
Likewise, for the ecological network, only some pollinators interact with many plants, while many pollinators only interact with a few plants. 
Following the literature about ecological networks, nodes that interact with a large number of nodes of the other type are called \emph{generalists} (G), while nodes that interact only with a small number of nodes of the other type are called \emph{specialists} (S).

In \emph{ecology}, the nested structure of bipartite networks has received much attention~\citep{Bascompte2007, Bascompte2003,rohr2014structural, Saavedra:2011, Atmar:1993, Memmott2004a, Mariani2019}.
In particular, the distinction between specialists and generalists was confirmed for many ecosystems.
It was shown~\citep{Bastolla:2009} that competition between species is minimised and biodiversity is increased when the incidence matrix has a nested structure.
Thus, the number of coexisting species in an ecological system is related to the nestedness of the incidence matrix.
It was argued~\citep{Suweis2013} that nestedness is a consequence of maximising the number of species in a mutualistic community, that is, of maximising an efficient resource usage.

In the same line, a  nested structure was also found for networks in \emph{economics}~\cite{May2008,Saavedra:2011,Garas2018}.
It was shown that the interactions of agents that compete for maximising their centrality within their local neighbourhood also leads to a nested network structure~\cite{THEC:THEC135}.
Since competition and efficient resource usage are driving forces behind the evolution of many natural, social and technological systems, it is reasonable to assume a  nested structure also for the inter-layer links in our multi-layer network.

\paragraph{Quantifying nestedness. \ }

We note that generalists and specialists are found among the nodes in \emph{both} layers.
A perfect nested structure implies a perfect ranking of generalists and specialists as illustrated in the incidence matrices $\mathbf{B}$ in Figure~\ref{fig:ex-arrangement}.
Real incidence matrices never show such a perfect nested structure, as Figure~\ref{fig:bio-bip}(b,d) indicates.

\begin{figure}[htbp]
  \centering
    \includegraphics[width=0.45\textwidth,angle=0]{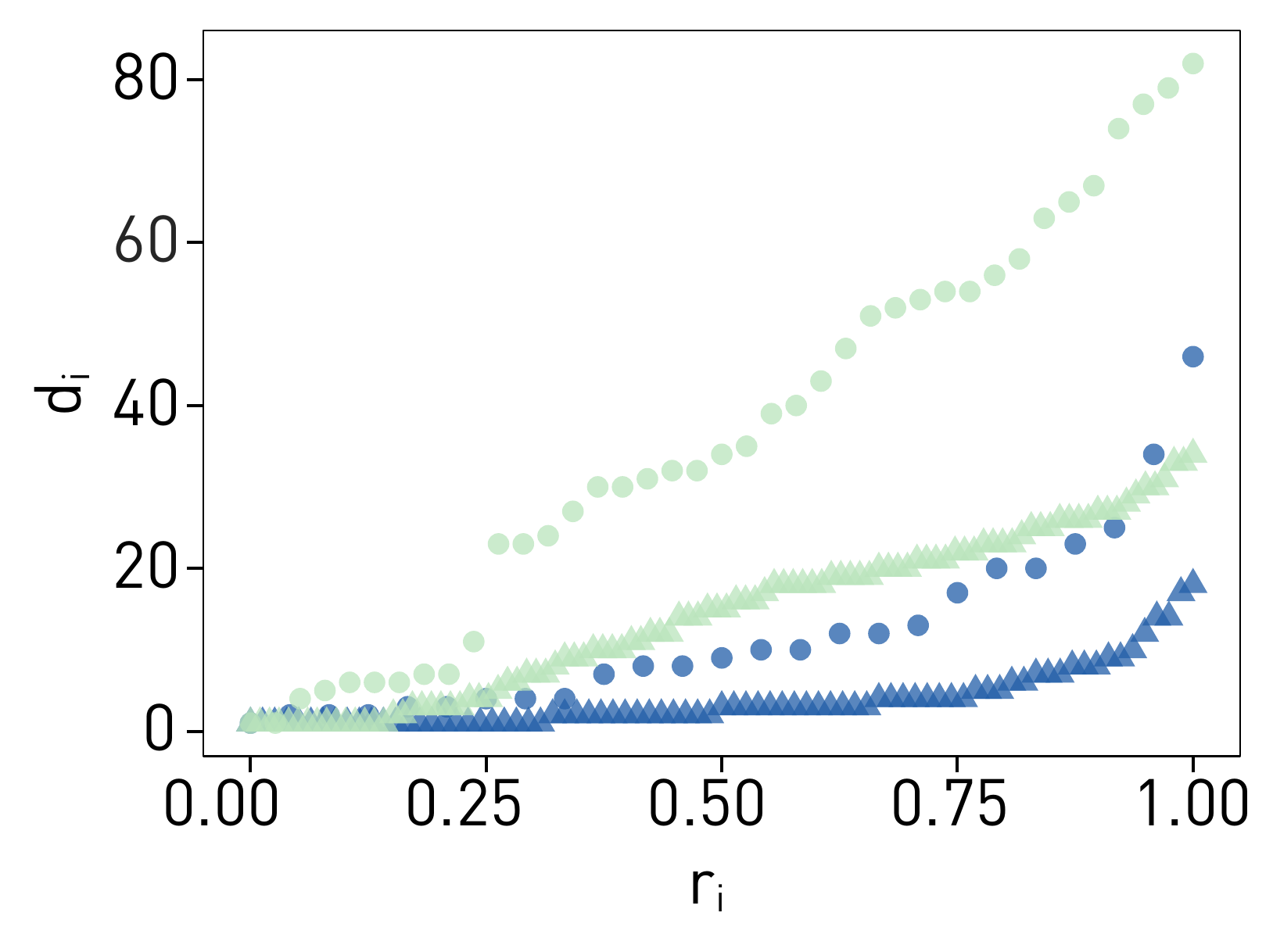}
  \caption[]{Ranking $r_{i}$ obtained from the NODF2 algorithm vs. inter-layer degrees $d_{i}$ for the two empirical networks shown in Figure~\ref{fig:bio-bip}(b,d).
  	Green denotes the economic network, blue the ecological network.
    Nodes in $\mathcal L_1$ are denoted by $\CIRCLE$, nodes in $\mathcal L_2$ are denoted by $\blacktriangle$.
    }
  \label{fig:ranking}
\end{figure}

In order to find the best possible ranking, an established procedure, the so-called NODF2 nestedness algorithm~\cite{Almeida-Neto:2008}, is used for sorting the incidence matrix.
This algorithm solves two problems: first, it provides a rank $r_{i}$ for each node that can be used to distinguish 
generalists and specialists.
We remark that the rank $r_{i}$ is \emph{not} a simple linear mapping of the  \emph{intra-layer degree}, $d_{i}$, although these two values are not independent. 
Figure~\ref{fig:ranking} shows the correlations between $r_{i}$ and $d_{i}$ for the respective layers of the two empirical bipartite networks.

Second, this algorithm returns a nestedness value $n_{\rm NODF}\ \in \ [0, 100]$ to characterise the whole bipartite network. 
$n_{\rm NODF} = 0$ when there is no nestedness and $n_{\rm NODF}=100$ for perfect nestedness.
We have applied NODF2 to the two empirical incidence matrices shown in  Figure~\ref{fig:bio-bip}(b,d), and obtained for the ecological network $n_{\rm NODF}=42.8$ and for the economic network $n_{\rm NODF}=75.9$ as the best possible values. 

To assess the significance of these numbers, we used a null model introduced in~\citep{Bascompte2003}.
It randomises the inter-layer links (who is connected to whom) but preserves the degree distribution (how many links a node has).
Specifically, we build an ensemble of bipartite networks that link nodes from layers 1 and 2 with a probability
proportional to their degree in the empirical bipartite network. 
Thus, the probability of node $i$ from $\mathcal{L}_{1}$ (e.g., a plant) to be linked to a node $j$ from $\mathcal{L}_{2}$ (e.g., a pollinator) is $p_{ij}=(d_i/N_1+d_j/N_2)/2$, where $d_i$, $d_{j}$ are the number of links of nodes $i$ and $j$ in the bipartite network.
$N_{1}$ and $N_{2}$ are the number of nodes in layers 1 and 2 of the bipartite network.
This way, we find, for example, that the nestedness value of $n_{\rm NODF}=75.9$ for the economic network is significant with $p<0.0001$.

We also use the NODF2 nestedness algorithm~\cite{Almeida-Neto:2008} to generate the structure of \emph{inter-layer} links in our multi-layer network. 
For this, we first choose a fill level, 35\%, which defines the density of the inter-layer links. 
Then, we arrange these links in a perfectly nested structure ($n_{\rm NODF}= 99.06$), shown in  the incidence matrices $\mathbf{B}$ in Figure~\ref{fig:ex-arrangement}.
Nodes from $\mathcal{L}_{1}$ with a high rank $r_{i}$ are \emph{generalists} in layer 1, denoted as $G_{1}$, while nodes from $\mathcal{L}_{1}$ with low rank are \emph{specialists}, denoted as $S_{1}$. 
Taking the perspective of layer 2, we can also indicate the generalists of $\mathcal{L}_{2}$, $G_{2}$, and the specialists of $\mathcal{L}_{2}$, $S_{2}$. 

With this, we have all the information together to characterise the nodes in the multi-layer network.
Each node $i\in N$ is assigned to a layer, either $\mathcal{L}_{1}$ or $\mathcal{L}_{2}$.
Its network position \emph{within} a layer is quantified by its coreness $k_{i}$, while its role in connecting the two layers is quantified by its rank $r_{i}$ which tells, with respect to a given layer, to what degree this node is a generalist or a specialist.
To say it again: \emph{core/periphery} always refers to a specific layer, while \emph{generalist/specialist} refers to the coupling between the two layers. 

\subsection{Scenarios to probe robustness}
\label{sec:scen-probe-robustn}

\paragraph{Attack scenarios. \ }

To probe the robustness of the multi-layer network, we utilise a classic approach~\citep{Albert2000}: (i) we remove an increasing fraction of nodes and their links from the network, (ii) we then calculate the size of the \emph{largest connected component} (LCC) of the remaining network~\citep{Schneider2011}.
The classic paper only considers \emph{one} network of a given topology, either a random graph or a scale-free network.
It then compares the impact on the LCC if randomly chosen nodes are removed or if nodes with high degree (hubs) are removed, specifically.
These different ways to choose nodes for removal were called \emph{attack scenarios}, a term we also use in the following. 
Not surprisingly, it was found that scale-free networks are robust against random attacks, but fragile if targeted attacks of hubs happen.
This finding was used to discuss the robustness of the internet, which can also be represented as a scale-free network.
But such a simplifying extrapolation was met with harsh criticism~\citep{Doyle2005}.
Different from the classic paper, in our model we have many more degrees of freedom to (i) consider attacks, and (ii) specify the multi-layer network, which needs to be done in the following.

\paragraph{Selection of nodes for removal. \ }

First, in addition to the \emph{degree} of nodes, which was the only characteristic used in the classic paper, we have information about the \emph{coreness}, which much better describes the embeddedness of nodes in a network.
Thus, we will choose the varying fraction of nodes to be removed with respect to their coreness values.
This makes a distinction between \emph{core} and \emph{periphery}, rather than between high and low degree nodes (hubs vs spokes). 

Second, we have a multi-layer network; therefore, in principle, we could remove nodes from different layers.
But for a more systematic approach, we decide to remove only nodes from \emph{one} layer, $\mathcal{L}_{1}$, to then study the effect on the \emph{other} layer, $\mathcal{L}_{2}$.
For this, we define the fraction of \emph{surviving} nodes in each layer as $F_{1}$, $F_{2}$.
Then $(1-F_{1})$ is the fraction of nodes that we \emph{remove} in $\mathcal{L}_{1}$.
To choose, which nodes shall be removed from $\mathcal{L}_{1}$, we consider 3 different \emph{attack scenarios}.
(1) nodes are \emph{randomly} chosen, i.e., with no particular order,
(2) nodes are chosen in the order of \emph{decreasing coreness} $k_{i}$ (i.e., we start with nodes from the \emph{core}),
(3) nodes are chosen in the order of \emph{increasing coreness} $k_{i}$ (i.e., we start with nodes from the \emph{periphery}). 
Because coreness values are degenerate, we choose randomly among the nodes with the same coreness value.

\paragraph{Removal of nodes. \ }

Specifically, the removal of nodes from $\mathcal{L}_{1}$ impacts $\mathcal{L}_{2}$ in the following way.
(i) A node from $\mathcal{L}_{1}$ is removed together with all its \emph{intra-layer} links and all its \emph{inter-layer} links.
(ii) The affected nodes $*$ in $\mathcal{L}_{2}$ therefore have less \emph{inter-layer} links, $d_{*}$. If $d_{*}>0$, these nodes are not removed and the procedure stops.
If $d_{*}=0$, i.e., if these nodes are no longer connected to $\mathcal{L}_{1}$, they are removed from $\mathcal{L}_{2}$ together with all their \emph{intra-layer} links.
(iii) If the removal procedure of links in $\mathcal{L}_{2}$ leaves other nodes isolated in $\mathcal{L}_{2}$, they will be removed as well together with all their \emph{inter-layer} links. 
The removal process stops when no nodes in $\mathcal L_2$ are left disconnected.

In other words, we assume that nodes in $\mathcal L_2$ depend on nodes both from $\mathcal L_1$ and $\mathcal L_2$.
The removal procedure in $\mathcal{L}_{1}$ results in a so-called \emph{secondary node loss} in $\mathcal L_2$, measured by $(1-F_{2})$, where $F_{2}$ is the fraction of surviving nodes in $\mathcal{L}_{2}$.
Because of the node and link removal in $\mathcal{L}_{2}$, there is a chance that the network fragments into disconnected components of various size (the procedure avoids isolated nodes).
Such a fragmented network can still have a large $F_{2}$ because that value does not reflect whether the network is connected or not.
This, on the other hand, would result in wrong conclusions about the robustness of the network, which is reduced because of the fragmentation. 
Therefore, we will calculate $F_{2}$ only on the \emph{largest connected component} of the network.
This ignores other components, but that the same time gives us a better insight into the  \emph{robustness} of the multi-layer network.

\paragraph{Influence of the inter-layer coupling. \ }

\begin{figure}[htbp]
\centering
\enskip\hfill\begin{subfigure}[b]{.22\textwidth}
  \includegraphics[angle=90,width=\textwidth]{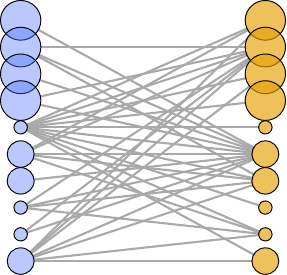}\\\caption{}\vspace{.5em}
  \includegraphics[width=\textwidth,angle=0]{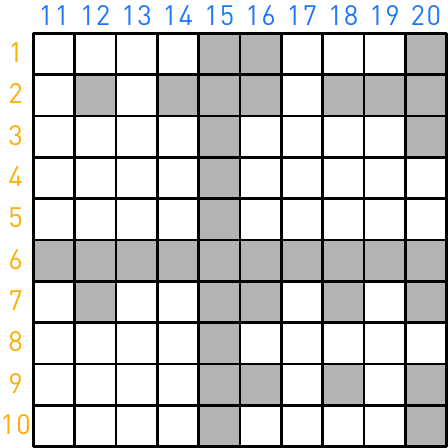}
\end{subfigure}\hfill
\quad\hfill\begin{subfigure}[b]{.22\textwidth}
  \includegraphics[angle=90,width=\textwidth]{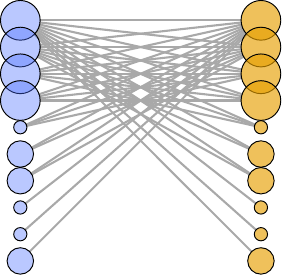}\\\caption{}\vspace{.5em}
  \includegraphics[width=\textwidth,angle=0]{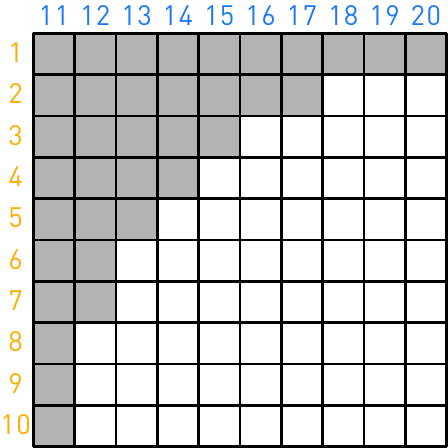}
\end{subfigure}\hfill
\quad\hfill\begin{subfigure}[b]{.22\textwidth}
  \includegraphics[angle=90,width=\textwidth]{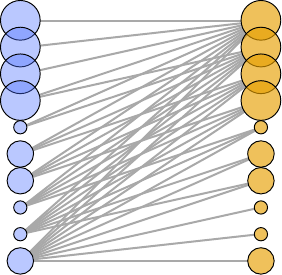}\\\caption{}\vspace{.5em}
  \includegraphics[width=\textwidth,angle=0]{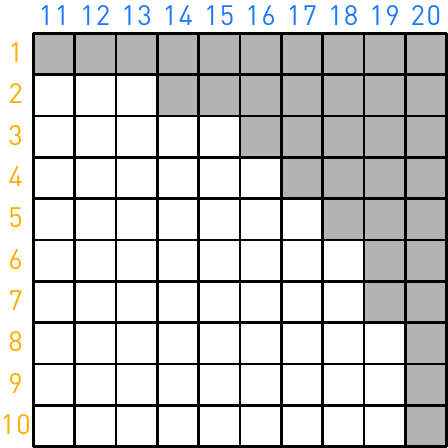}
\end{subfigure}\hfill
\quad\hfill\begin{subfigure}[b]{.22\textwidth}
  \includegraphics[angle=90,width=\textwidth]{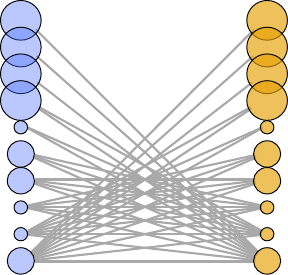}\\\caption{}\vspace{.5em}
  \includegraphics[width=\textwidth,angle=0]{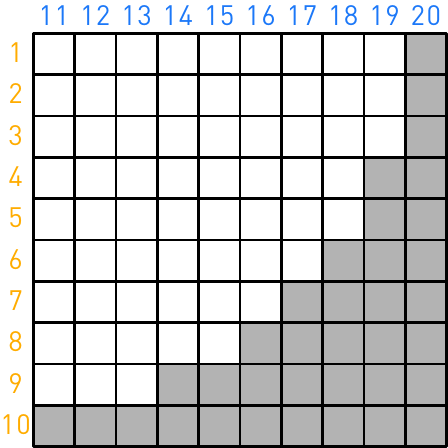}
\end{subfigure}\hfill\enskip

  \caption[]{Different couplings in a two-layer network of $N=20$ nodes:  
    {\bf (a)} random connection, used as a null model;
    {\bf (b)} cG$_{1}$-cG$_{2}$;
  {\bf (c)} cG$_{1}$-cS$_{2}$;
  {\bf (d)} cS$_{1}$-cS$_{2}$.
  The nodes in both layers are ranked in decreasing order of $r_{i}$ with respect to (b) (i.e., generalists on the left, specialists on the right).
  The size of the nodes is proportional to their coreness $k_{i}$, obtained in a scale-free core-periphery network.
For all couplings, the incidence matrices $\mathbf{B}$ are also shown.}
  \label{fig:ex-arrangement}
\end{figure}

As the last degree of freedom in our modelling approach, we have to specify further how the two layers are coupled.
We recall that, in each layer, nodes are characterised by their \emph{inter-layer} degree $d_{i}$ and by their ranking $r_{i}$, to reflect their role as generalists or specialists.
Using this information, we can discuss different ways of inter-layer coupling between generalists and specialists in each layer.
In the following, we focus on \emph{core nodes}, i.e., nodes characterised by a high coreness value $k_{i}$ in their respective layer.
In Figure~\ref{fig:ex-arrangement}, the size of the nodes is chosen proportional to their coreness value (because the network is small, some nodes have only small $k_{i}$).

Figure~\ref{fig:ex-arrangement} illustrates different ways to couple generalists and specialists: 
(a)  a random connection between core nodes from $\mathcal{L}_{1}$ and $\mathcal{L}_{2}$, which we use as a reference case,
(b) cG$_{1}$-cG$_{2}$, i.e., core nodes from $\mathcal{L}_{1}$ that are generalists are coupled to core nodes from $\mathcal{L}_{2}$ that are also generalists,
(c)  cG$_{1}$-cS$_{2}$, i.e., core nodes from $\mathcal{L}_{1}$ that are generalists are coupled to core nodes from $\mathcal{L}_{2}$ that are specialists,
(d)   cS$_{1}$-cS$_{2}$, i.e., core nodes from $\mathcal{L}_{1}$ that are specialists are coupled to core nodes from $\mathcal{L}_{2}$ that are also specialists.
For each combination, we also show the respective incidence matrix $\mathbf{B}$.

\section{Results}
\label{result}

\subsection{Robustness of the ecological sample network}
\label{sec:robustn-memmott}

To get a better idea how to interpret the results of our probing procedure, we start with a real-world example, the \emph{ecological} bipartite network between plant and pollinator species shown in Figure~\ref{fig:bio-bip}(a,b).
This network is different from the multi-layer networks discussed in Section~\ref{sec:full-model} in that it \emph{only} contains \emph{inter-layer} links and \emph{no} intra-layer links.
Therefore, we cannot apply our attack scenarios (1)-(3), to remove nodes based on their \emph{coreness} values, $k_{i}$.
Instead, we take the only information available and remove nodes according to their intra-layer degrees, $d_{i}$.
Because there are no intra-layer links and, hence, no components, $F_{2}$ gives the fraction of all nodes surviving in $\mathcal{L}_{2}$ 
Thus, we get a feeling of how the robustness of the multi-layer network is affected \emph{in the absense} of intra-layer links, but with a \emph{nested structure} to couple both layers.

\begin{figure}[htbp]
\centering
\quad\hfill\begin{subfigure}[b]{0.39\textwidth}
 \includegraphics[width=\textwidth]{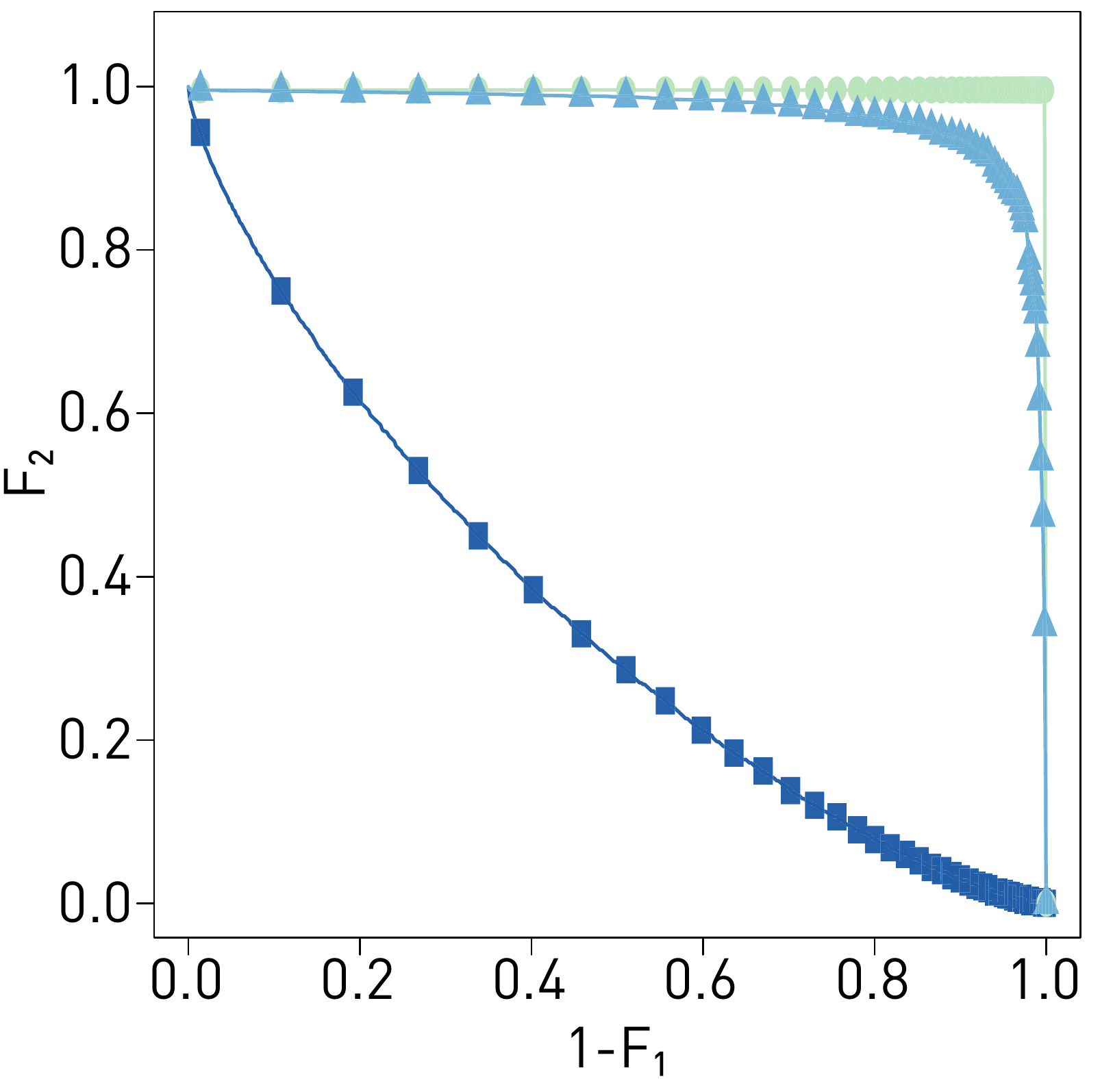}
 \caption{}
\end{subfigure}\hfill
\quad\hfill\begin{subfigure}[b]{0.49\textwidth}
  \includegraphics[width=\textwidth]{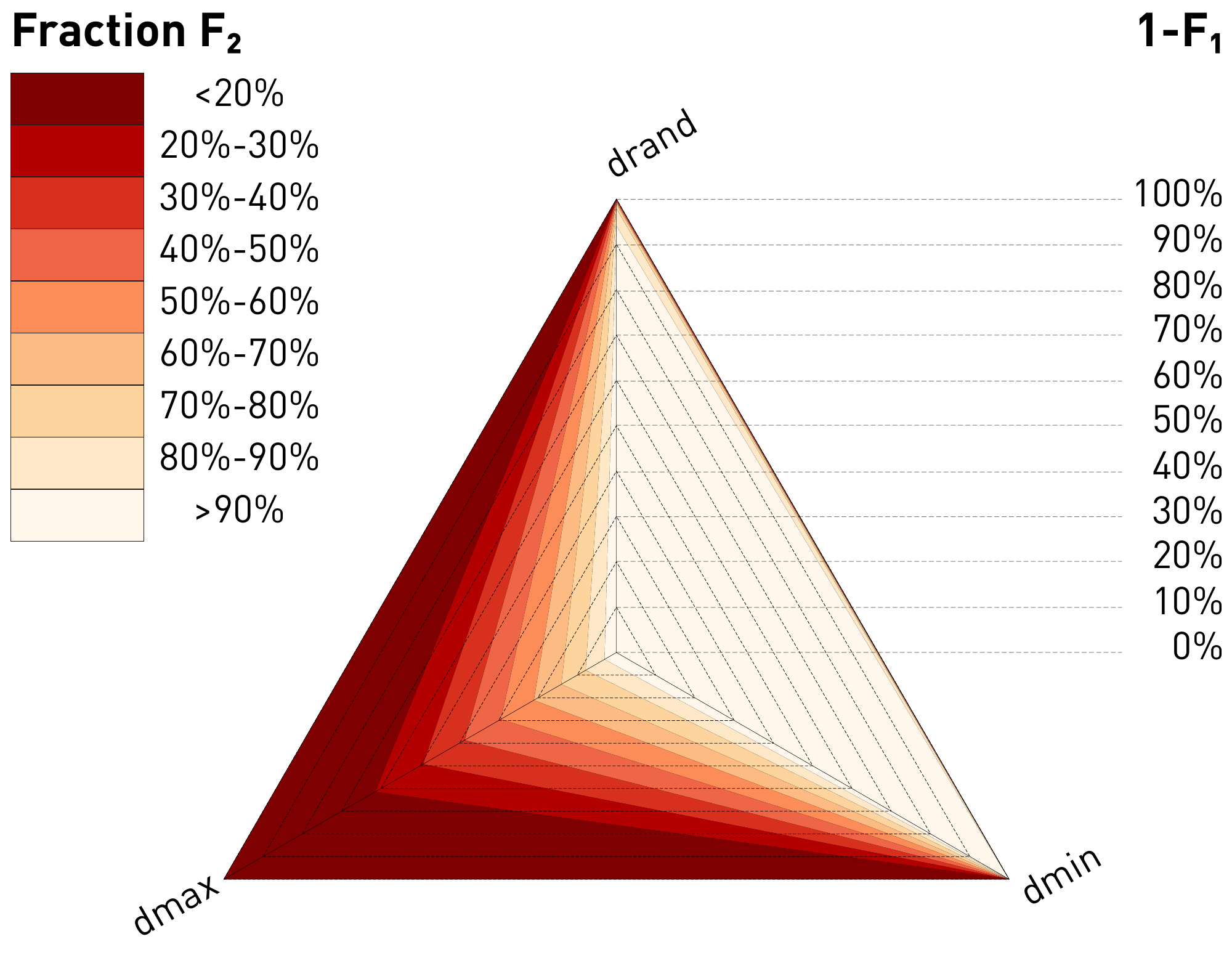}
  \caption{}
  \end{subfigure}\hfill
\quad
  \caption[]
  {\textbf{(a)} Fraction of nodes surviving in $\mathcal{L}_{2}$ dependent on the fraction of nodes removed from $\mathcal{L}_{1}$. 
    $\blacktriangle$: random removal scenario,  $\blacksquare$: nodes are removed in the order of decreasing $d_{i}$,
    $\CIRCLE$: nodes are removed in the order of increasing $d_{i}$.
      The results are averaged over 100 simulations, and the standard errors are smaller than the size of the symbols.
\textbf{(b)} Robustness profile.
}
  \label{fig:ex-memmott}
\end{figure}

The data of the ecological network comes from~\citep{Robertson1929}.
The robustness of this nested network was already studied in~\citep{Memmott2004a}.
There, very similar to our approach, different attack scenarios for node removal were applied to the pollinators in $\mathcal{L}_{1}$, to measure the impact on the plants in $\mathcal{L}_{2}$.
We reproduce the respective results in Figure~\ref{fig:ex-memmott}(a), to also verify that our method leads to the same outcome as shown in~\citep{Memmott2004a} (there: Figure 1b).
In Figure~\ref{fig:ex-memmott}(b) we show the robustness profile as a different way of visualising the results. 

Figure~\ref{fig:ex-memmott}(a) shows that the ecological network is very robust against the \emph{random} removal of pollinator species.
Only if more than 80\% are removed, a noticeable impact on the plant species can be observed, and at about 95\% a breakdown happens.
The situation is even better if instead of a random removal, removal of nodes in the order of increasing $d_{i}$ is considered.
Nodes with small $d_{i}$ are very likely peripheral nodes in the bipartite network and at the same time very likely specialists (see Figure~\ref{fig:ranking}a). 
Thus, the network is extremely robust against their removal -- as long as generalists are still available.
If, however, the removal starts with nodes of high $d_{i}$, which are likely core nodes and generalists in the bipartite network (see Figure~\ref{fig:ranking}),
the robustness decreases very quickly and faster than linear with the fraction of removed nodes.
This reminds on the fragility of scale-free networks in case hubs are removed, but, different from the synthetic networks discussed afterwards, the ecological network studied here is not a scale-free network. 

The robustness profile in Figure~\ref{fig:ex-memmott}(b) compacts these insights.
The three axes of the spider-plot refer to the three different attack scenarios.
On each axis, the fraction of removed nodes from $\mathcal{L}_{1}$ increases from the origin towards the corner.
The colour codes the robustness of the network, as measured by $F_{2}$: the lighter, the more robust the network.
Deep red colour thus indicates an outcome where less than 20\% of nodes in $\mathcal{L}_{2}$ have survived.
It is obvious that this happens if nodes with high $d_{i}$ are chosen first to be removed.

With respect to the ecological network, we can conclude that systems characterised by mutualistic interactions, which are usually described by nested structures of their incidence matrix, are particularly robust~\cite{Morris2003,Memmott2004a}.

\subsection{Robustness of the full model}
\label{sec:full-model}

With this information, we now turn to the full model of multi-layer networks that also include \emph{intra-layer} links, as shown in Figure~\ref{fig:exmulti-3}.
We remove nodes from $\mathcal{L}_{1}$ according to their coreness values, $k_{i}$, as described in the different attack scenarios (1)-(3) in Section~\ref{sec:scen-probe-robustn}, and we measure the impact on $\mathcal{L}_{2}$ by the fraction of the largest connected component of surviving nodes, $F_{2}$. 
In addition to the attack scenarios, we also vary the inter-layer coupling, as shown in Figure~\ref{fig:ex-arrangement}.
Our results are obtained from scale-free two-layer networks with 1000 nodes and averaged over 100 simulations.
The standard errors are smaller than the size of the symbols.

\begin{figure}[htbp]
\centering
\quad\hfill\begin{subfigure}[b]{0.36\textwidth}
 \includegraphics[width=\textwidth]{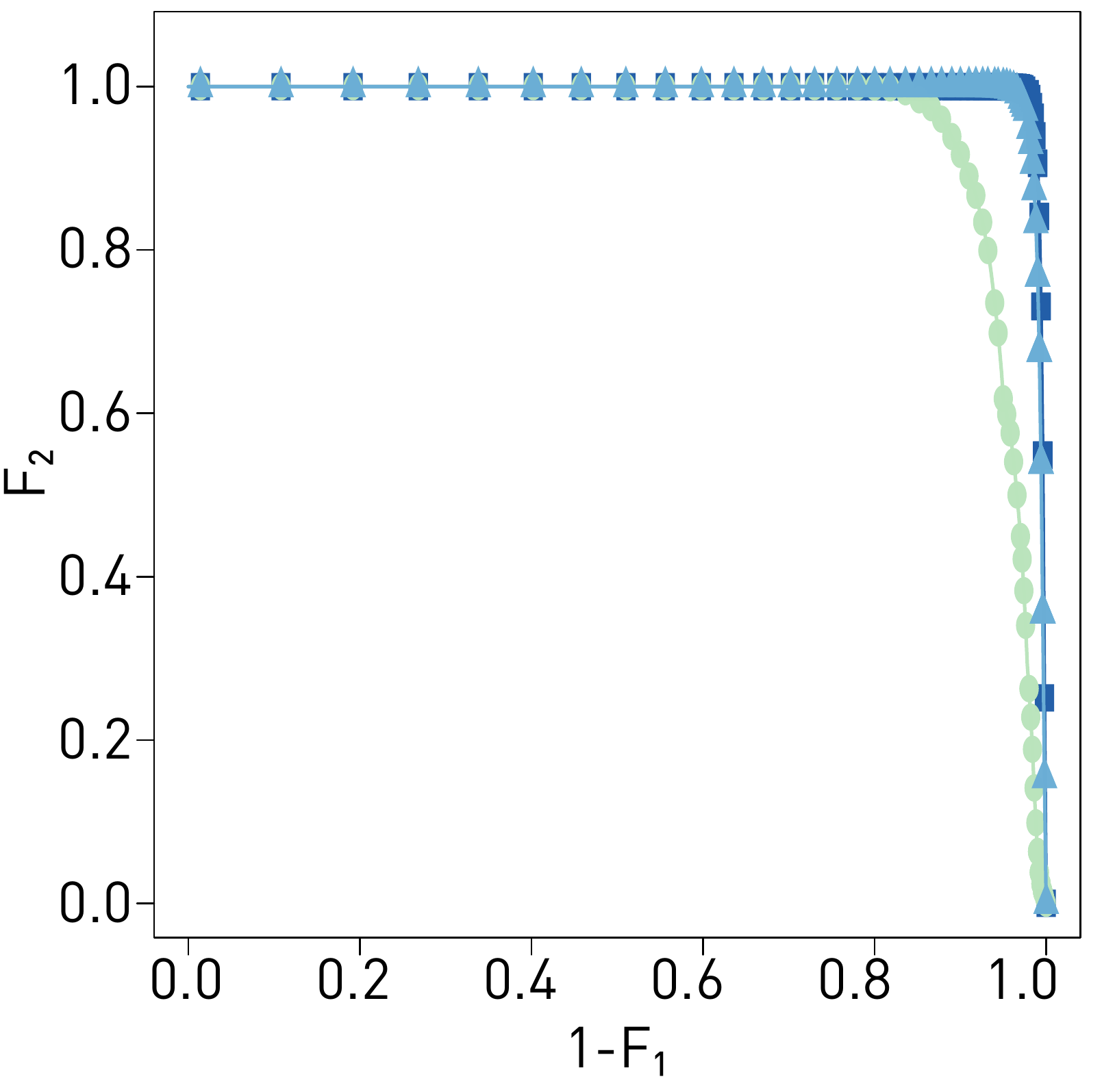}
 \caption{}
\end{subfigure}\hfill
\quad\hfill\begin{subfigure}[b]{0.46\textwidth}
  \includegraphics[width=\textwidth]{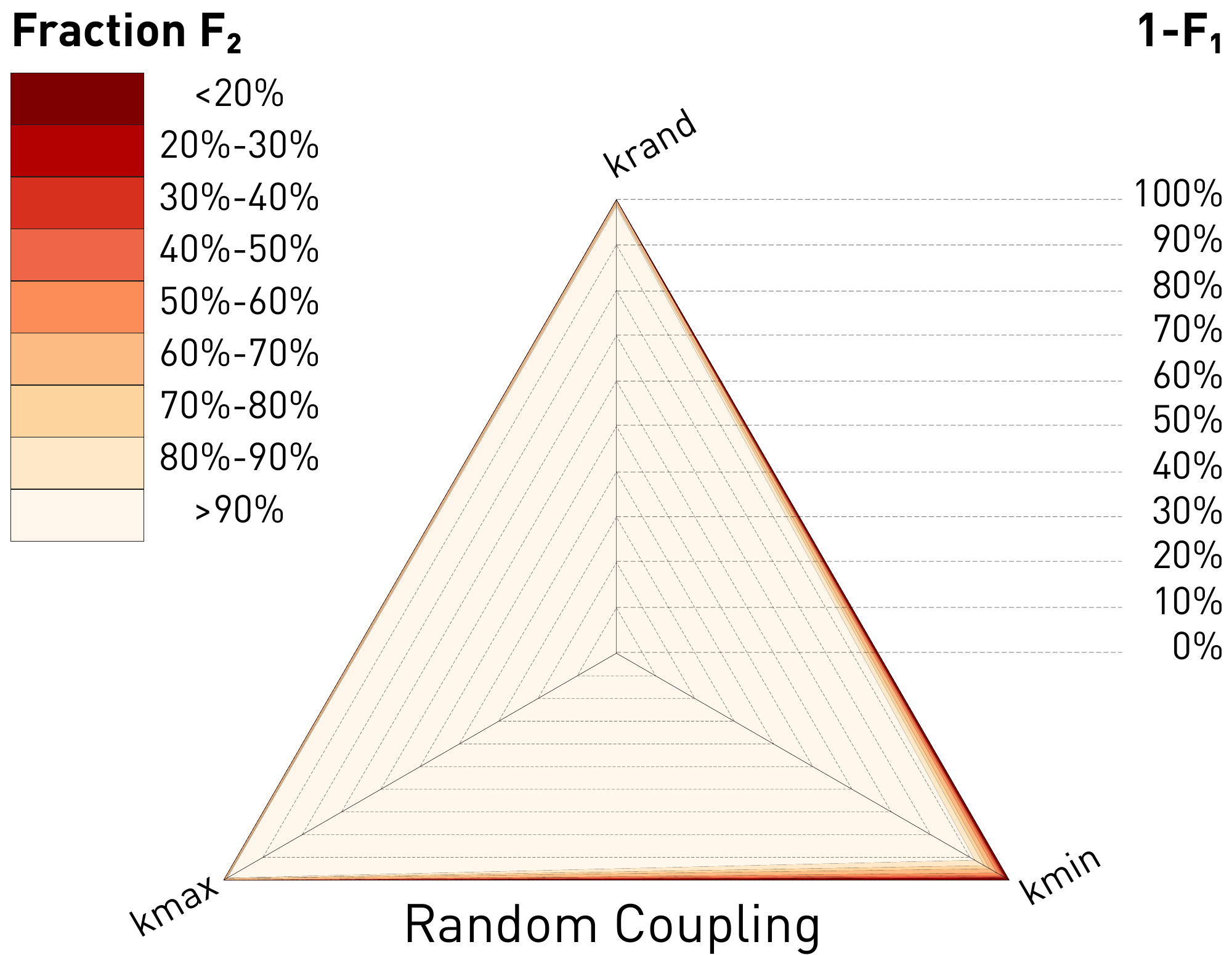}
  \caption{}
  \end{subfigure}\hfill\quad
  \caption[]
  {Random coupling between $\mathcal{L}_{1}$ and $\mathcal{L}_{2}$. \textbf{(a)} Fraction of nodes surviving in $\mathcal{L}_{2}$ dependent on the fraction of nodes removed from $\mathcal{L}_{1}$. 
    Attack scenarios: (1) $\blacktriangle$,  (2) $\blacksquare$, (3) $\CIRCLE$. 
\textbf{(b)} Robustness profile: $k_{\mathrm{rand}}$ refers to attack scenario (1), $k_{\mathrm{max}}$ to (2), $k_{\mathrm{min}}$ to (3). 
}
  \label{fig:random}
\end{figure}

Let us first discuss a reference case, namely the random coupling between layers, Figure~\ref{fig:ex-arrangement}(a), and the random removal of nodes, attack scenario (1).
The results are shown in Figure~\ref{fig:random}.
We verify that scale-free networks are robust against random node removal also for the case of two-layer networks with intra- and inter-layer links.
This insight does not change if instead of random removal nodes with decreasing coreness are removed (2), and it only slightly worsened if instead the removal of nodes with increasing coreness is considered (3).
We note that in this reference case, due to the random coupling between the two layers, the nested structure is destroyed, as also shown in Figure~\ref{fig:ranking-GS}(a).

\begin{figure}[htbp]
\centering\vspace{-2.5em}
\quad\hfill\begin{subfigure}[b]{0.35\textwidth}
 \includegraphics[width=\textwidth]{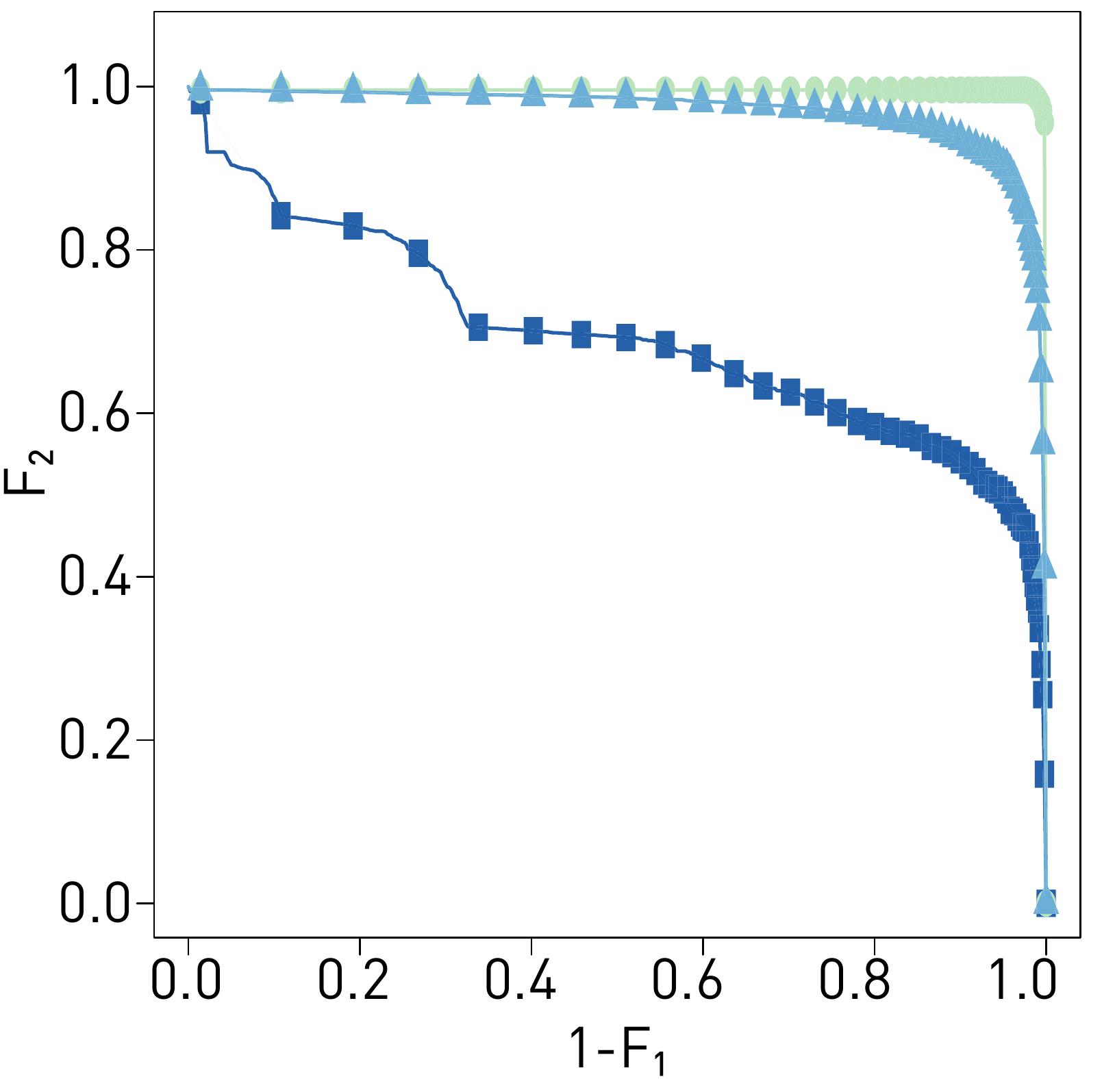}
 \caption{}
\end{subfigure}\hfill
\quad\hfill\begin{subfigure}[b]{0.45\textwidth}
  \includegraphics[width=\textwidth]{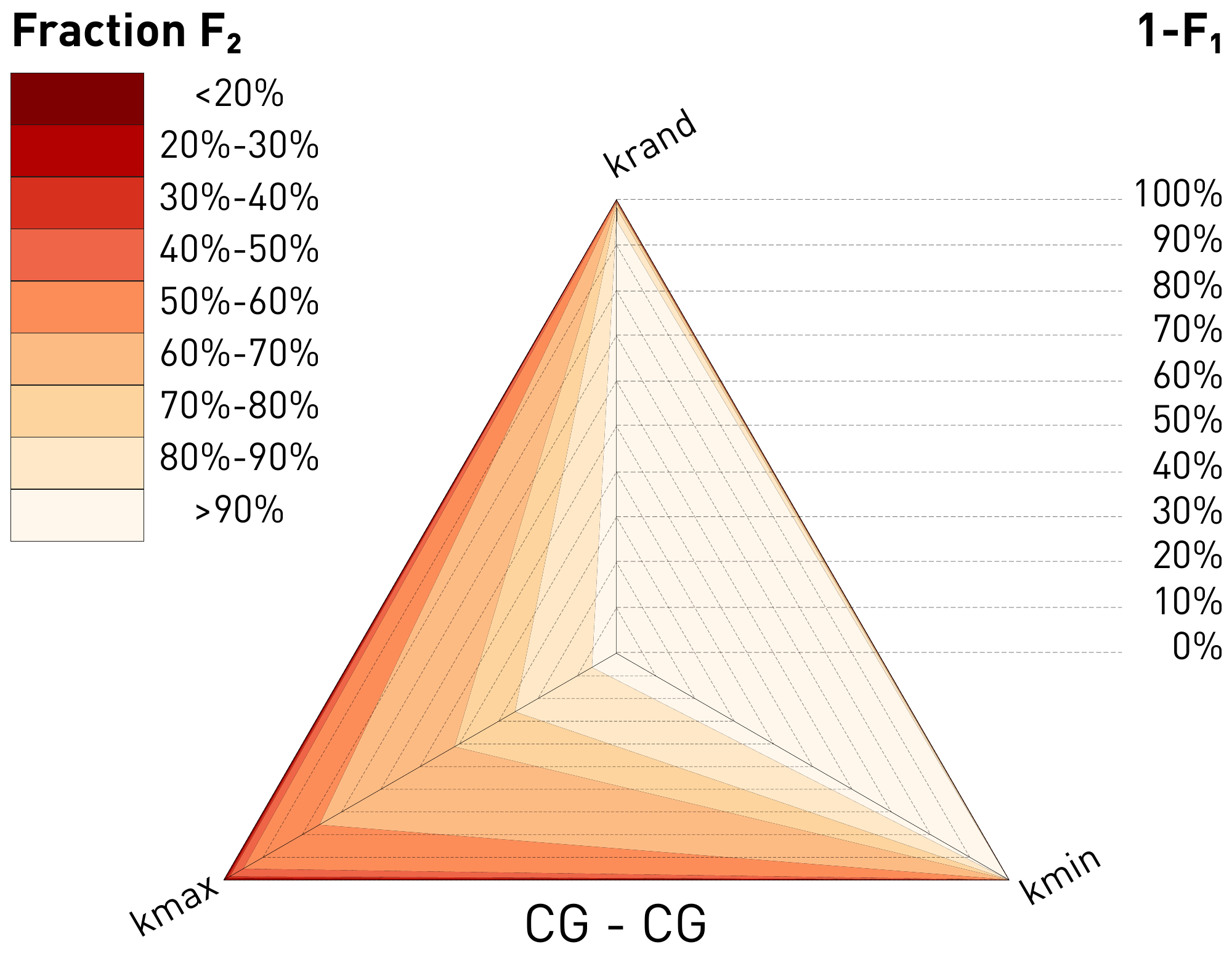}
  \caption{}
  \end{subfigure}\hfill\quad

\vspace{.5em}
\quad\hfill\begin{subfigure}[b]{0.35\textwidth}
 \includegraphics[width=\textwidth]{./figures/LCC_L2_g4}
 \caption{}
\end{subfigure}\hfill
\quad\hfill\begin{subfigure}[b]{0.45\textwidth}
  \includegraphics[width=\textwidth]{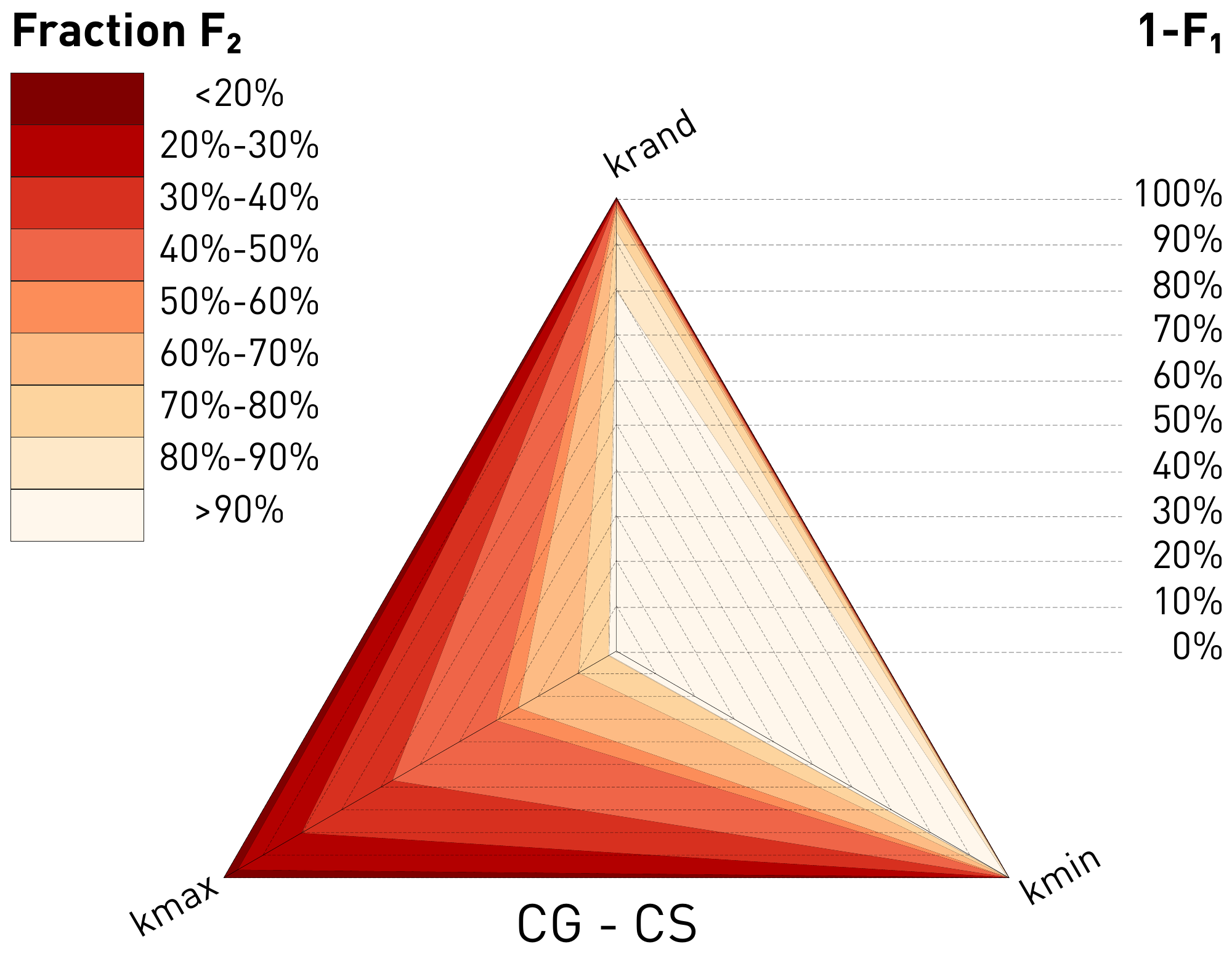}
  \caption{}
  \end{subfigure}\hfill\quad

\vspace{.5em}
\quad\hfill\begin{subfigure}[b]{0.35\textwidth}
 \includegraphics[width=\textwidth]{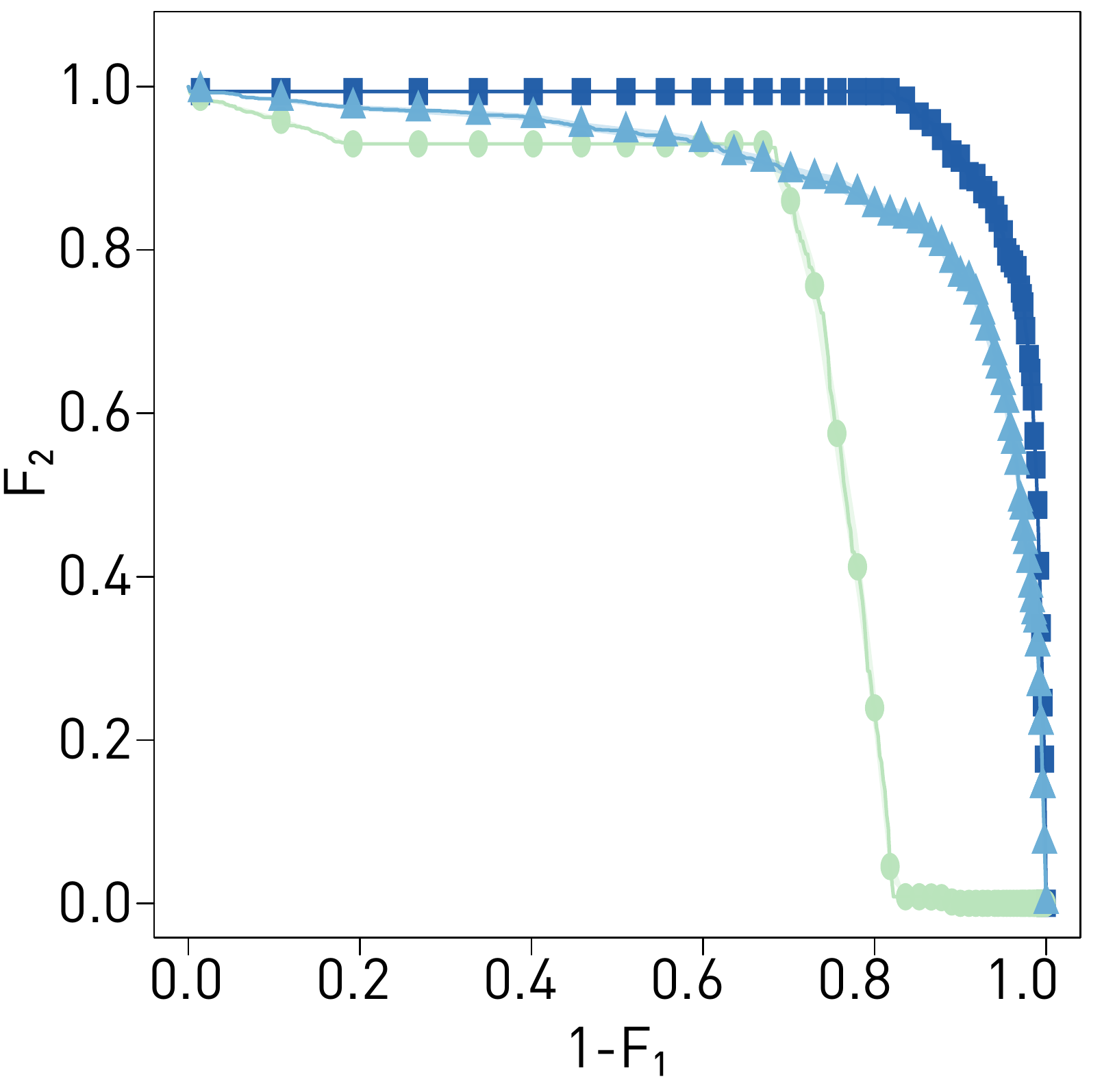}
 \caption{}
\end{subfigure}\hfill
\quad\hfill\begin{subfigure}[b]{0.45\textwidth}
  \includegraphics[width=\textwidth]{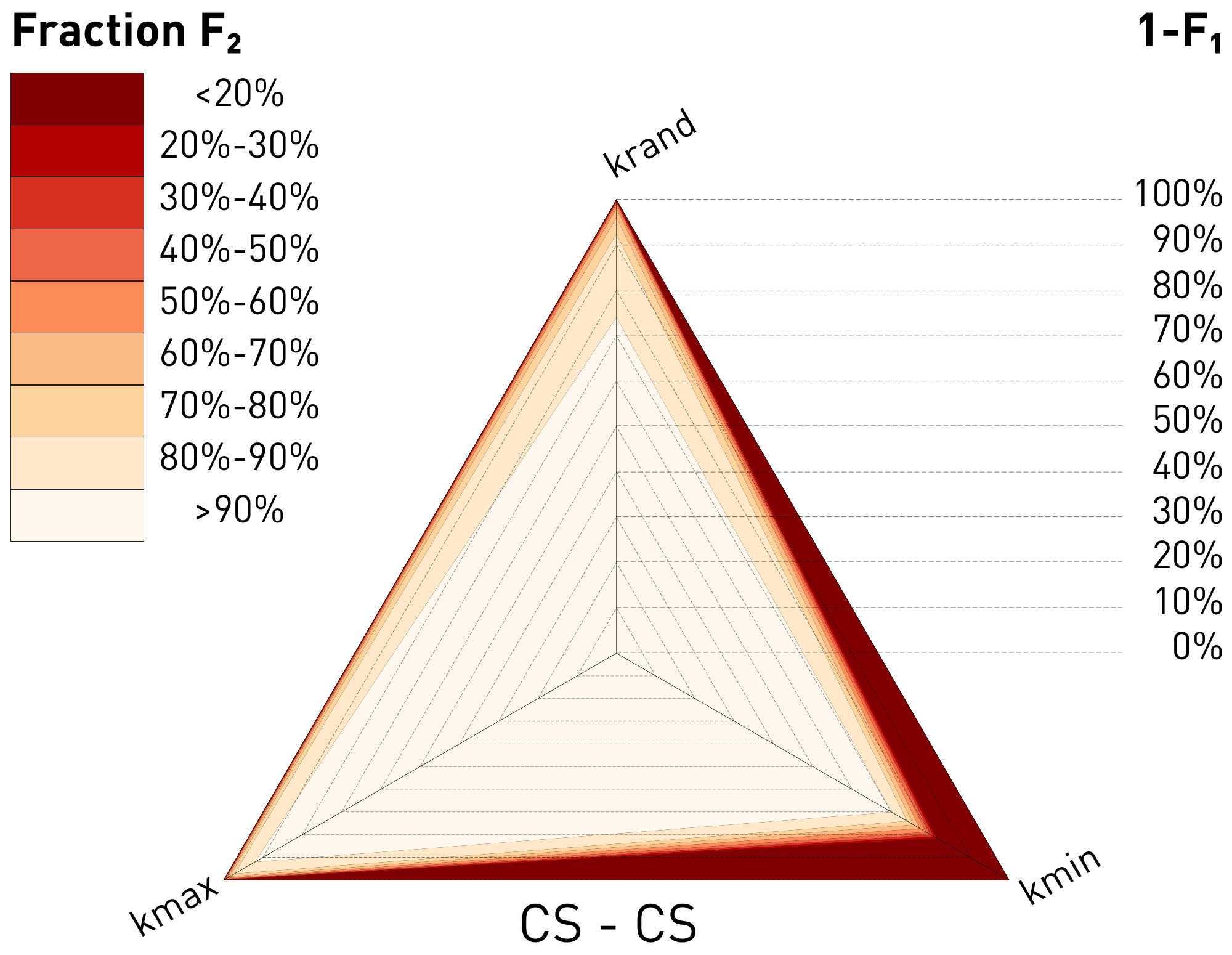}
  \caption{}
  \end{subfigure}\hfill\quad
  \caption[]
  {\textbf{(left)} Fraction of nodes surviving in $\mathcal{L}_{2}$ dependent on the fraction of nodes removed from $\mathcal{L}_{1}$. Coupling between $\mathcal{L}_{1}$ and $\mathcal{L}_{2}$: (a,b) cG$_{1}$-cG$_{2}$, (c,d) cG$_{1}$-cS$_{2}$, (e,f) cS$_{1}$-cS$_{2}$. 
        Attack scenarios: (1) $\blacktriangle$,  (2) $\blacksquare$, (3) $\CIRCLE$. 
\textbf{(right)} Robustness profiles:  $k_{\mathrm{rand}}$ refers to attack scenario (1), $k_{\mathrm{max}}$ to (2), $k_{\mathrm{min}}$ to (3). 
}
  \label{fig:together}
\end{figure}

To further investigate the role of nestedness, we now discuss the three different couplings that respect this nested structure.
The results are shown together in Figure~\ref{fig:together}.
We start with coupling generalists in both layers, shown in Figure~\ref{fig:together}(a,b).
We find that the removal of core nodes from $\mathcal{L}_{1}$, (2), has a considerable impact on the robustness, whereas the other two attack scenarios do not decrease the robustness.
Precisely, the removal of the 10\% top core nodes in $\mathcal L_1$ can lead to almost 20\% secondary node losses in $\mathcal L_2$.
This bears similarities with the plots for the ecological network, Figure~\ref{fig:ex-memmott}, although the decrease of the respective curve (2) is not as quick for the multi-layer network.

If we now connect the generalists of $\mathcal{L}_{1}$ with the specialists of $\mathcal{L}_{2}$ (instead of the generalists), we find that things get worse when core nodes from $\mathcal{L}_{1}$ are removed (2).
As shown in Figure~\ref{fig:together}(c,d), in this case the removal of the 10\% top core nodes in $\mathcal L_1$ leads to almost 40\% secondary node losses in $\mathcal L_2$, while a removal of the 40\% top core nodes in $\mathcal L_1$ totally destroys the network in $\mathcal L_2$, already. 
This is a noticeable difference to the other two attack scenarios, which have no (1) or little (3) impact on the robustness of the multi-layer network.

If, on the other hand, we connect the specialists of both layers, we find that the removal of core nodes from $\mathcal{L}_{1}$ (2) has very little impact, as Figure~\ref{fig:together}(e,f) shows.   
It requires the systematic removal of more than 80\% of the core nodes in $\mathcal L_1$ to trigger any secondary node loss in $\mathcal L_2$.
We see instead that the multi-layer network becomes less robust against the removal of peripheral nodes from $\mathcal{L}_{1}$ (3).
Also, the random removal (1) results in a larger decrease of robustness than the removal of top core nodes (3).
For removal of peripheral nodes up to 70\%, the scenarios (1) and (3) give the same results, which are \emph{both} worse than for scenario (3).

\begin{figure}[htbp]
\centering
\begin{subfigure}[b]{0.45\textwidth}
 \includegraphics[width=\textwidth]{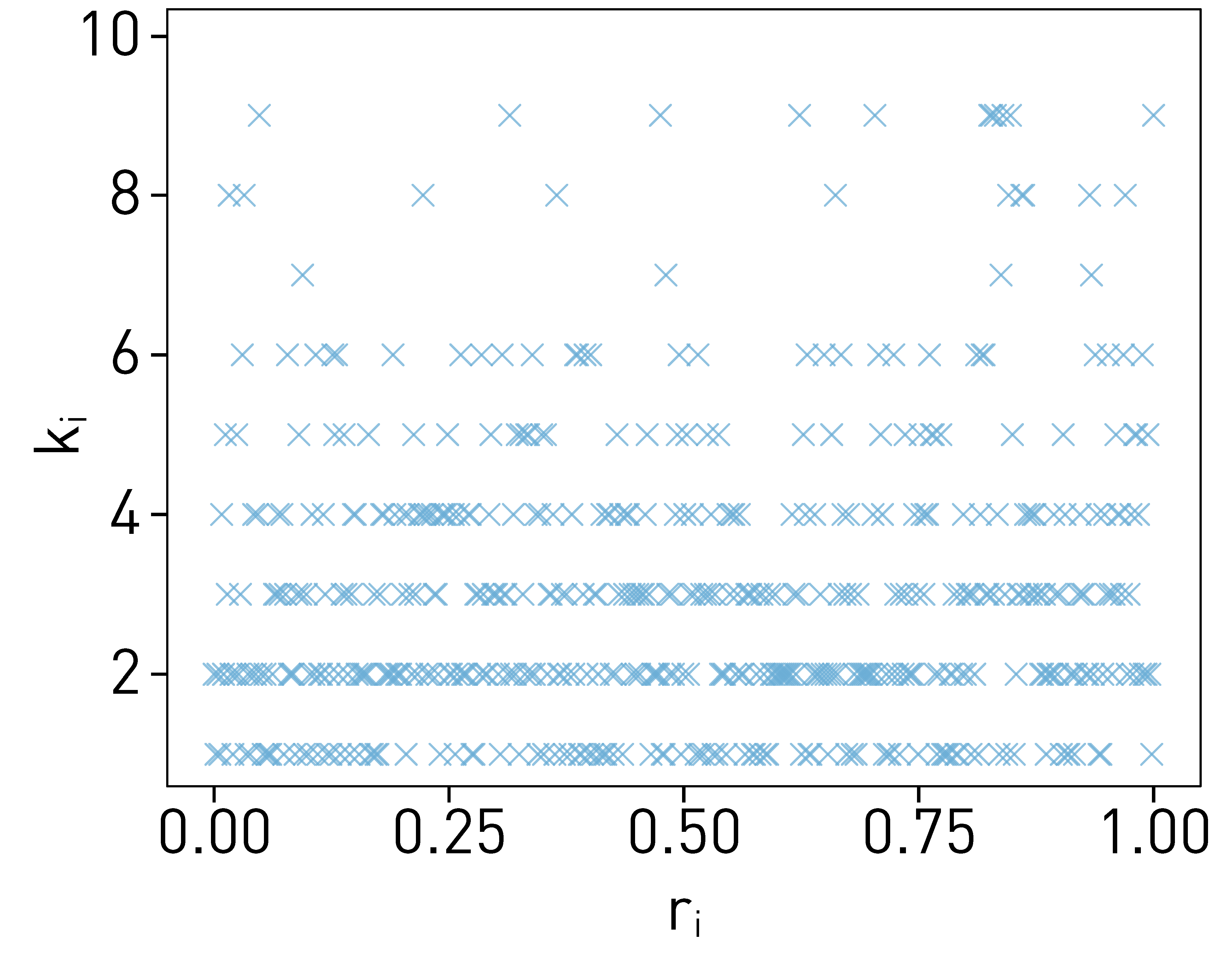}
 \caption{}
\end{subfigure}\hfill
\begin{subfigure}[b]{0.45\textwidth}
  \includegraphics[width=\textwidth]{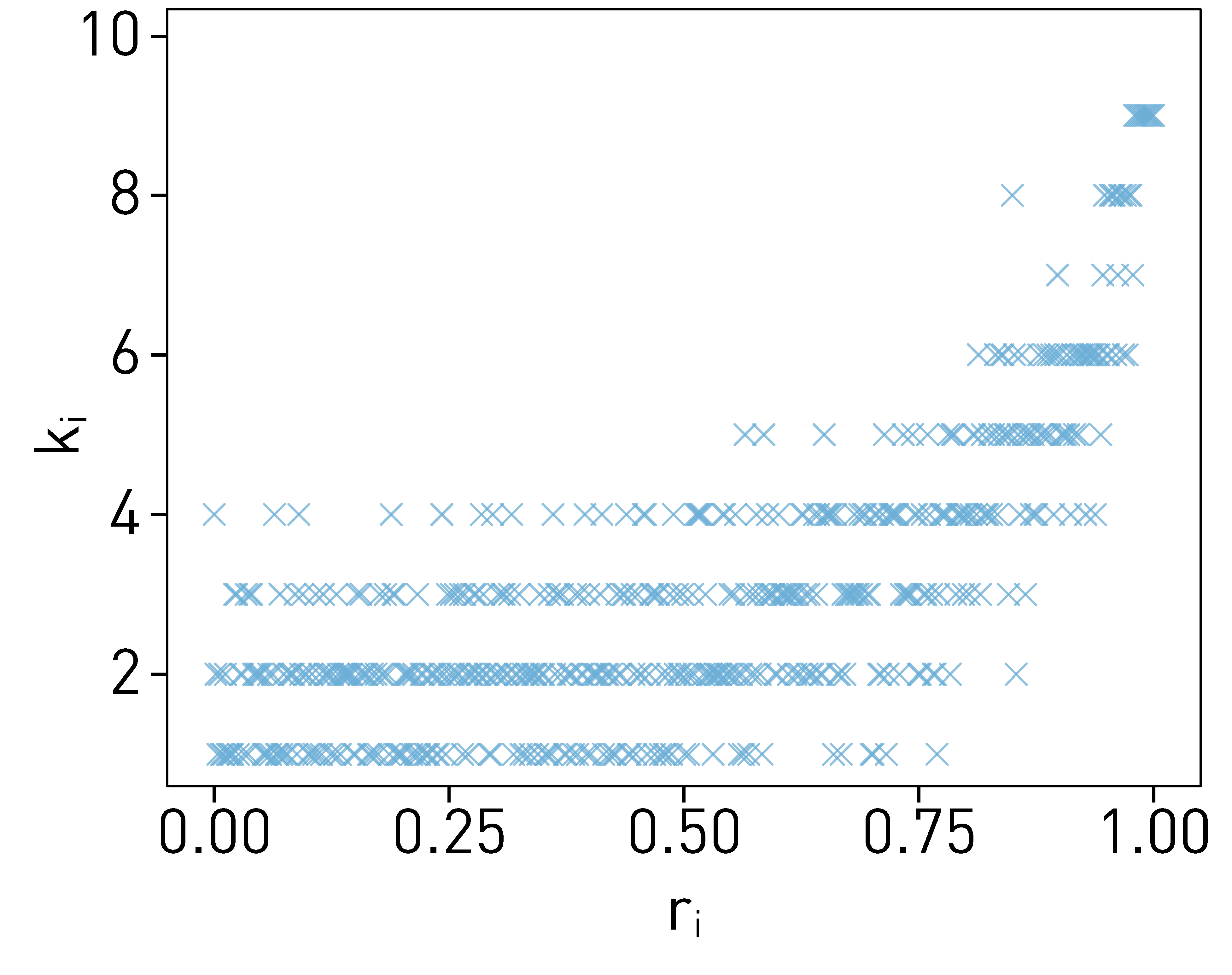}
  \caption{}
  \end{subfigure}
  \caption[]{Ranking $r_{i}$ obtained from the NODF2 algorithm for the 500 nodes of $\mathcal{L}_{1}$ vs. coreness values $k_{i}$, considering (a) random coupling, (b) coupling of core generalists (cG$_{1}$).
    We note that the ranking obtained for the coupling of core specialists (cS$_{1}$) would just invert the $x$ axis in (b). 
    }
  \label{fig:ranking-GS}
\end{figure}

In conclusion, we find that the impact of the different attack strategies (1)-(3) strongly depends on the inter-layer coupling.
When the \emph{generalists} of $\mathcal{L}_{1}$ are coupled to $\mathcal{L}_{2}$ (either to the generalists or the specialists), we find that removing \emph{core} nodes from $\mathcal{L}_{1}$ leads to the strongest decrease on the robustness of the multi-layer network.
On the other hand, if the \emph{specialists} of $\mathcal{L}_{1}$ are coupled to the generalists of $\mathcal{L}_{2}$, we find that the removal of \emph{peripheral} nodes from $\mathcal{L}_{1}$ reduces the robustness more.

This insight cannot be simply explained by the argument that generalists are often core nodes, and specialists are often peripheral nodes.
As Figure~\ref{fig:ranking-GS}(b) shows, for scale-free networks the correlations between coreness $k_{i}$ and ranking $r_{i}$ are not   that simple -- not even in the case of \emph{perfect nestedness}, we considered here. 
Nodes with low coreness, i.e., \emph{peripheral} nodes, can still have considerably high ranks, i.e., play the role of \emph{generalists}.
Nodes with high coreness, i.e., \emph{core} nodes, on the other hand, can have ranks comparable to peripheral nodes.
Therefore, the interplay between inter-layer coupling and attack strategies need to be taken into account to explain the robustness of the multi-layer network.

\section{Discussion}\label{discussion}

In this paper, we have extended the discussion about the robustness of scale-free networks to multi-layer networks with different couplings between the layers.
For the topology within and across layers, we have utilised insights from empirical networks.
For the \emph{intra-layer} topology, we considered a \emph{core-periphery} structure, which was quantified by assigning a \emph{coreness} value to each node.
For the \emph{inter-layer} topology, we considered a \emph{nested} structure, which was quantified by assigning a \emph{rank} to each node, obtained from the NODF2 algorithm.

The \emph{robustness} of the multi-layer network was probed by removing a fraction of nodes from layer $\mathcal{L}_{1}$ together with their intra- and inter-layer links, and measuring the fraction of nodes surviving in the LCC of layer $\mathcal{L}_{2}$.
We assumed that nodes in $\mathcal{L}_{2}$ which became disconnected from $\mathcal{L}_{1}$ because of the node removal are removed in $\mathcal{L}_{2}$ together with their intra-layer links.

In our approach we varied two model features: (i) the attack scenarios within $\mathcal{L}_{1}$: removing nodes either randomly (1), with decreasing (2) or increasing coreness (3), and (ii) the coupling of generalists and specialists between $\mathcal{L}_{1}$ and $\mathcal{L}_{2}$: either randomly, or cG$_{1}$-cG$_{2}$, cG$_{1}$-cS$_{2}$, cS$_{1}$-cS$_{2}$.

We found that for the robustness of multi-layer networks, the two features need to be considered together.
While it is already known that nested bipartite networks are particularly robust~\cite{Morris2003,Memmott2004a}, 
the nested structure alone is not sufficient to explain the robustness.
Instead, the intra-layer topology, specifically the core-periphery structure, also plays an important role.
Removing core nodes first does not always lead to a larger decrease in robustness, and removing peripheral nodes first does not always lead to a smaller decrease.
The impact of these attack scenarios depends on whether generalists or specialists are coupled.
The strongest effect was observed for the coupling cG$_{1}$-cS$_{2}$ and the removal of core nodes.
Already 40\% of nodes removed from $\mathcal{L}_{1}$ was sufficient to destroy the multi-layer network.
This outcome could not be simply anticipated, but it results from the non-linear impact of the two features combined in this specific manner.

Our theoretical findings are of relevance for ecological systems, which are very often only studied as bipartite networks, i.e., neglecting \emph{intra-layer} links.
We found that considering such links may eventually lead to higher robustness for certain nested structures.
Or, to put it the other way round, reducing multi-layer networks, for example, ecological or economic ones,  to a bipartite network structure, results in wrong estimations about their robustness.
This points to the availability of empirical data to reconstruct such multi-layer networks instead of bipartite ones.
Such data is rare, but 
in principle, there are ways to infer interactions in ecological systems from limited information~\citep{Kefi2015}.

With respect to modelling the robustness of multi-layer networks, which was our main focus in this paper, we see two different extensions of the scope. 
The first one regards a more refined modelling of the impact of node removals in $\mathcal{L}_{1}$.
We have considered that the intra-layer and inter-layer links of these nodes are removed as well.
But we have not considered that the removal of individual nodes generates \emph{failure cascades}, because other nodes in $\mathcal{L}_{1}$ may have been disconnected.
This is a major difference to investigations about \emph{systemic risk}, which reflect that the failure (or the removal) of a few nodes may be amplified into large failure cascades.

To model this amplification, we need more information about the vulnerability of the nodes, but also about their interaction with neighbouring nodes.
Systemic risk is then quantified as the fraction of failed nodes at the end of a failure cascade.
There exist already mathematical frameworks to calculate systemic risk for single layer~\citep{Gleeson2008,Burkholz2018} and multi-layer~\citep{Burkholz2016} networks.
For the latter case, it was shown that a critical coupling strength between the two layers exists.
Below the critical value, failure cascades can be reduced, and the multi-layer network is more robust, above the critical value.
However, these cascades are even amplified between the layers, and the multi-layer network becomes less robust.

The second extension regards the \emph{control} of multi-layer networks such that their robustness is increased.
For single-layer networks, a framework for network controllability was proposed in~\citep{Liu2011}.
It identifies nodes that are most influential in driving the network towards the desired state.
Applying this framework, however, requires assumptions about a linear dynamics that runs on the network to allow for the control.
Extensions to the control of multi-layer networks have been proposed already~\citep{bianconi2018multilayer,ZhangACS}.
Interestingly, it was demonstrated for different inter-layer couplings that peripheral nodes are as valuable as core nodes in controlling multi-layer scale-free networks~\citep{Zhang2016}. 

Such network interventions are a promising alternative to improve the robustness of multi-layer networks.
But, again, to apply the respective frameworks requires much more information about the state of the nodes and their interactions.
In our paper, we have chosen only a minimal set of assumptions, to have a clear focus on the impact of different inter-layer couplings -- the main difference between single-layer and multi-layer networks.

\small \setlength{\bibsep}{1pt}

\end{document}